\documentclass[useAMS,usenatbib] {mn2e}
\usepackage{graphicx}
\usepackage{times}
\usepackage{natbib}

\begin{document}

\title[Cosmological Shocks: Main Properties and CR]{Shock Waves in 
Eulerian Cosmological Simulations: Main Properties and 
Acceleration of Cosmic Rays}
\author[F. Vazza, G. Brunetti, C. Gheller]{F.Vazza$^{1,2}$\thanks{
E-mail: vazza@ira.inaf.it}, G. Brunetti$^{2}$, C. Gheller$^{3}$ \\
$^{1}$ Dipartimento di Astronomia, Universit\'a di Bologna, via Ranzani
1,I-40127 Bologna, Italy \\
$^{2}$ INAF/Istituto di Radioastronomia, via Gobetti 101, I-40129 Bologna,
Italy\\
$^{3}$ CINECA, High Performance System Division, Casalecchio di
Reno--Bologna, Italy}
\date{Accepted ???. Received ???; in original form ???}
\maketitle

\begin{abstract}

Large Scale Shocks are responsible for
the heating of the ICM and can be
important sources of Cosmic Rays (CR) in the Universe.
However the occurrence and properties of these
shocks are still poorly constrained from both a theoretical
and an observational side.

In this work we analyze the properties of Large Scale Shocks
in a $(103 {\rm Mpc/h})^{3}$ cosmological volume simulated
with the public 1.0.1 release of the ENZO code. 
Different methods to identify and characterize shocks in post processing
are discussed together with their uncertainties. 
Re-ionization affects the properties of shocks in 
simulations, and we propose a 
fitting procedure to model accurately the 
effect of re-ionization in non--radiative simulations, 
with a post--processing procedure.
We investigate the 
properties of shocks in our simulations by means of a 
procedure which uses jumps in the
velocity variables across the cells in the simulations
and this allows us to have a viable description of
shocks also in relatively under-dense cosmic regions. 
In particular 
we derive the distributions of the number of shocks
and of the energy dissipated at these shocks as a function
of their Mach number,
and discuss the evolution of these distributions
with cosmological time and across different cosmic environments
(clusters, outskirts, filaments, voids).

In line with previous numerical studies (e.g. Ryu et al.2003, Pfrommer et al.2006) relatively weak shocks
are found to dominate the process of energy dissipation in the simulated
cosmic volume, although
we find a larger ratio between weak and strong shocks with respect
to previous studies.
The bulk of energy is dissipated at shocks with Mach number
$M \approx 2$ and the fraction of strong shocks decreases
with increasing the density of the cosmic
environments, in agreement with semi-analytical studies in the case 
of galaxy clusters.

We estimate the rate of injection of CR at Large 
Scale Shocks by adopting injection efficiencies taken from previous
numerical calculations. 
The bulk of the energy is dissipated   
in galaxy clusters and in filaments and the
flux dissipated in the form of CR within the whole
simulated volume at the present epoch is $\approx 0.2$ of the thermal
energy dissipated at shocks; this fraction is smaller 
within galaxy clusters.

Finally we discuss the properties of shocks in 
galaxy clusters in relation with their dynamical state.
In these regions the bulk of the energy is dissipated at weak
shocks, with Mach number $\approx 1.5$, although slightly stronger
shocks are found in the external regions of merging clusters.
\end{abstract}


\label{firstpage} 

\vskip 0.4cm \fontsize{11pt}{11pt} \selectfont  

\section{Introduction}

\label{sec:intro}

\bigskip

Galaxy clusters store up to several $10^{63}$ ergs in the form of hot
baryonic matter, with typical temperatures of several keV. This thermal 
energy is mostly the by-product of shock-heating processes occurred 
during the formation of
cosmic structures.
However detecting shocks in Large Scale Structures (LSS) is still 
observationally challenging since they usually
develop in the external regions of galaxy clusters, where 
the X--ray emission is faint.
In a few cases, however, internal shocks driven by the merging events have been
discovered with typical  
Mach numbers $\approx 1.5 - 3$ (e.g. Markevitch et al. 1999; 
Markevitch et al. 2002; Belsole et al.2004; Markevitch \& Vikhlinin 2007).
Studies concerning the
convolution of simulated cosmological shocks into simulated X--ray images
(e.g. Gardini et al. 2004; Rasia et al. 2006) have clearly shown that the
apparent lack of shocks in the ICM may be due to the concourse of
projections and mass--weighting effects along the line of sight, 
which tend to bias
towards lower values of the overall temperature of the ICM and to smooth 
temperature gradients.

Shocks are important not only to understand the heating of the ICM 
but also because they may be efficient accelerators of 
supra--thermal particles (e.g. Sarazin 1999; 
Takizawa \& Naito 2000; Blasi 2001).
Non thermal activity in galaxy clusters is
proved by radio observations
which show synchrotron emission in a fraction of merging clusters:
Radio Halos , at the cluster center, and 
Radio Relics, elongated sources at the cluster periphery (e.g.,
Feretti 2005). Several mechanisms related to 
cluster mergers and to the accretion of matter can act as sources of 
non thermal components in galaxy clusters.
Large scale radio emission in the form of 
giant Radio Halos may be powered by particle re-acceleration 
by MHD turbulence injected
in the ICM during energetic merging events (Brunetti et al. 2001; 
Petrosian 2001; Brunetti et al. 2004,08).
Strong shocks may accelerate supra-- thermal electrons 
from the thermal pool and explain the origin of 
Radio Relics (Ensslin et al.1998), 
while high energy electrons accelerated at these shocks can produce
X-rays and gamma-rays via Inverse Compton scattering off CMB photons (e.g.
Sarazin 1999; Loeb \& Waxmann 2000; Blasi 2001; Miniati 2003).
Relativistic hadrons accelerated at shocks can be advected in
galaxy clusters and efficiently 
accumulated (V{\"o}lk, Aharonian, \& 
Breitschwerdt 1996; Berezinsky, Blasi, \& 
Ptuskin 1997) producing an
important non-thermal component which could be directly sampled
by future gamma ray observations (e.g., Blasi, Gabici \& Brunetti 2007).
Secondary particles injected in the ICM via proton--proton
collisions may also produce detectable synchrotron radiation
(e.g. Blasi \& Colafrancesco 1999; Dolag \& Enssil 2000) 
and may be eventually re-accelerated by MHD turbulence yielding an efficient
picture to explain Radio Haloes (Brunetti \& Blasi 2005).

The energetics associated with the population of cosmic ray particles (CR)
accelerated at shocks depends on the Mach number of these shocks 
(e.g. Kang \& Jones 2002).
The Mach number distribution of cosmological shocks is thus
important to understand CR in galaxy clusters.
Semi--analytical studies pointed out that shocks that form 
during cluster mergers are weak, $M \sim 1.5$, being driven by sub-clumps 
crossing the main clusters at the free-fall velocity (Gabici \& Blasi 2003,
Berrington \& Dermer 2003).
These approaches however are limited as they treat cluster mergers 
as binary encounters between ideally virialised spherical systems.
Therefore cosmological numerical simulations represent a necessary 
avenue to address this issue in more detail.
First attempts to characterize shock waves in cosmological
numerical simulations were produced by Miniati et al.(2000),
by employing a set of eulerian simulations and a shock detecting
scheme based on jumps in the temperature variable.
Later works adopted more refined shock-detecting schemes and
were more focused onto the distribution of
energy dissipated at shocks (Keshet et al.2003; Ryu et al.2003, 
Hallman et al.2003, Pfrommer et al.2006{\footnote{After this paper was
submitted, also Skillman et al.2008 provided results on shock waves in ENZO AMR simulations, which are broadly consistent with the above works in literature.}}).
Ryu et al. and Pfrommer et al. basically found that the
bulk of shocks in the universe is made 
of relatively weak shocks, but also
allow to constrain the population 
of stronger shocks forming in the external 
regions of galaxy clusters, were structures are not completely virialised.
In these environments, strong shocks are frequent and may 
provide the bulk of the acceleration of CR in large scale
structures (Ryu et al. 2003; Pfrommer et al. 2006).
For this reason numerical simulations are important to study 
the non-thermal components in galaxy clusters
(e.g. Miniati et al.2001, Miniati 2003; Keshet et al.2003; Pfrommer 2008).
However, the identification 
and characterization of shocks, 
as well as the calculation of the energy injected 
in the form of CR at these shocks, is difficult because of
the complex dynamics of large scale structures and because of the 
severe limitations in terms of physics and numerical resolution that affect
present cosmological simulations.

In this paper we follow the approach of the seminal paper
by Ryu et al. (2003), studying the shock wave patterns 
in LSS and characterizing them in a post--processing
phase. This allows us to briefly discuss the effect of different
shock detection techniques and their dependence
on the variation of underlying physical processes 
(e.g. re-ionization) in the simulation.
We use the cosmological code ENZO (Bryan \& Norman 1997),
which treats gas dynamics with an Eulerian scheme and
allows us to follow the assembly of LSS with sufficiently good
spatial resolution.

The outline of the paper is the following.
In Sect.\ref{sec:enzo} we provide a brief introduction to the ENZO Code,
in Sect.\ref{sec:sim} we present our cluster sample and the
main properties of cosmological
structures in our simulations, and in Sect.\ref{sec:reionization} we discuss the effect of
re-ionization on the thermal properties of simulated cosmic structures.
In Sect.\ref{sec:algo} we provide the different methods to characterize shocks 
in post processing and in Sect.\ref{sec:comparison} we discuss their
main source of uncertainties in the cosmological framework.
In Sect.\ref{sec:results} we present our results about
the main shocks properties and about the
injection of CR.
The main conclusions of this work are given in Sect.\ref{sec:conclusions}.
In the Appendix we the effect of the resolution and of $\sigma_{8}$ 
on our results.

\begin{table}
\label{tab:tab0}
\caption{Main characteristics of the simulations.}
\centering \tabcolsep 4pt 
\begin{tabular}{c|c|c|c}
 Volume &  Resolution &  physics & ID\\ \hline
 $(145Mpc)^{3}$ & $125kpc$ & adiab. & {\bf AD125}\\
 $(145Mpc)^{3}$ & $250kpc$ & adiab. & {\bf AD250}\\
 $(145Mpc)^{3}$ & $500kpc$ & adiab. & {\bf AD500}\\
 $(145Mpc)^{3}$ & $800kpc$ & adiab. & {\bf AD800}\\
 $(80Mpc)^{3}$ & $125kpc$  & cool. + reion. & {\bf CO125}\\
 $(80Mpc)^{3}$ & $250kpc$ & cool., reion. and $\sigma8=0.74$ & {\bf S8250}\\
 $(80Mpc)^{3}$ & $250kpc$ & cool.+reion. & {\bf CO250}\\
\end{tabular}
\end{table}

\section{Numerical Code - ENZO.}
\label{sec:enzo}

A precise description of the behavior of the bayonic gas is crucial for 
the goals of the present work. In particular, the numerical 
code adopted for our simulations must support an accurate 
treatment of the dynamics of high supersonic 
flows and the
formation and propagation of strong shock waves during the process of
cosmological structures formation.
The ENZO code supports such description.
ENZO is an adaptive mesh refinement (AMR) cosmological hybrid 
originally written by Greg Bryan and Michael Norman
(Bryan \& Norman 1997, 1998; Norman \& Bryan 1999; Bryan, Abel, \& 
Norman 2001, Norman et al.2007). It couples
an N-body particle-mesh solver with an adaptive mesh method for ideal 
fluidynamics (Berger \& Colella, 1989).
ENZO adopts an Eulerian hydrodynamical solver
based on the the Piecewise Parabolic 
Method (PPM, Woodward \& Colella, 1984). 
The PPM algorithm belongs to a class
of schemes in which an accurate representation of flow discontinuities is made
possible by building into the numerical method the calculation of the
propagation and interaction of non--linear waves. 
It is a higher order extension of Godunov's shock capturing
method (Godunov 1959). It is at least second--order accurate in space (up
to the fourth--order, in the case of smooth flows and small time-steps) and
second--order accurate in time. 
The PPM method describes shocks with high accuracy 
and has no need of artificial viscosity, leading to an 
optimal treatment of energy
conversion processes, to the
minimization of errors due to the finite size of the cells of the grid and
to a spatial resolution close to the nominal one. In the cosmological
framework, the basic PPM technique has been modified to include the
gravitational interaction and the expansion of the universe.
For a detailed review of the ENZO
code and a comparison with other numerical approaches,  
we refer to O'Shea et al.(2004) and O'Shea et al.(2005). 

In this work, in order to keep our study of LSS shocks as simple 
as possible, we use ENZO with a fixed spatial resolution without the 
application of AMR techniques.

\begin{figure}
\includegraphics[width=0.45\textwidth]{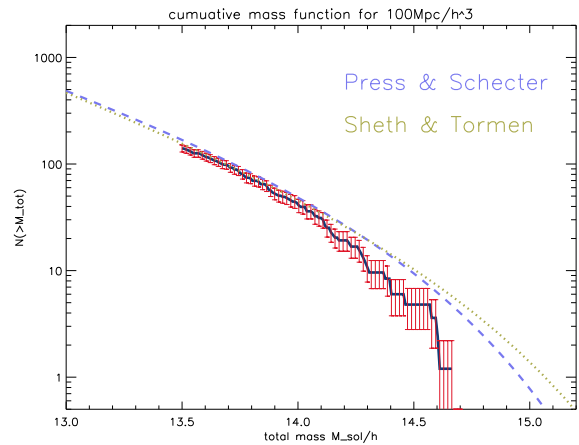}
\caption{Cumulative Mass Function for the total matter of all halos in the simulations,
with poissonian
errors. Press \& Schechter ({\it dashed}) and Sheth \& Tormen ({\it dotted}) 
mass functions
are reported for comparison.}   
\label{fig:massfunc}
\end{figure}

\begin{figure}
\includegraphics[width=0.45\textwidth]{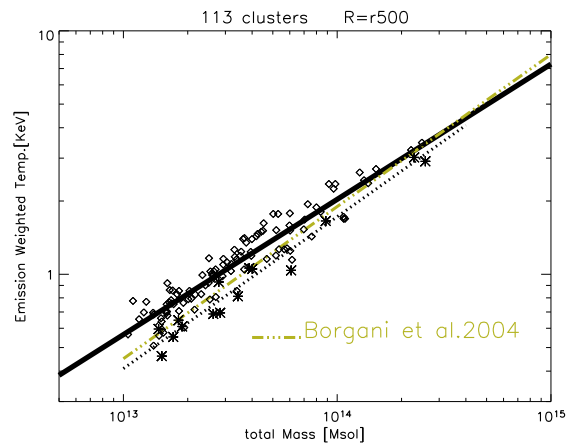}
\caption{Total Mass versus Emission--Weighted Temperature for the
simulated galaxy clusters of our {\it AD125} run (diamonds). 
Points for 15 clusters 
of the {\it CO125} run are also plotted (asterisks). Best fit relations 
for these samples
are drawn ({\it solid line = AD125, dashed = CO125}), together with 
a comparison 
the results of Borgani et al.2004 ({\it yellow line}).}
\label{fig:M_T}
\end{figure}

\begin{figure*}
\includegraphics[width=0.95\textwidth]{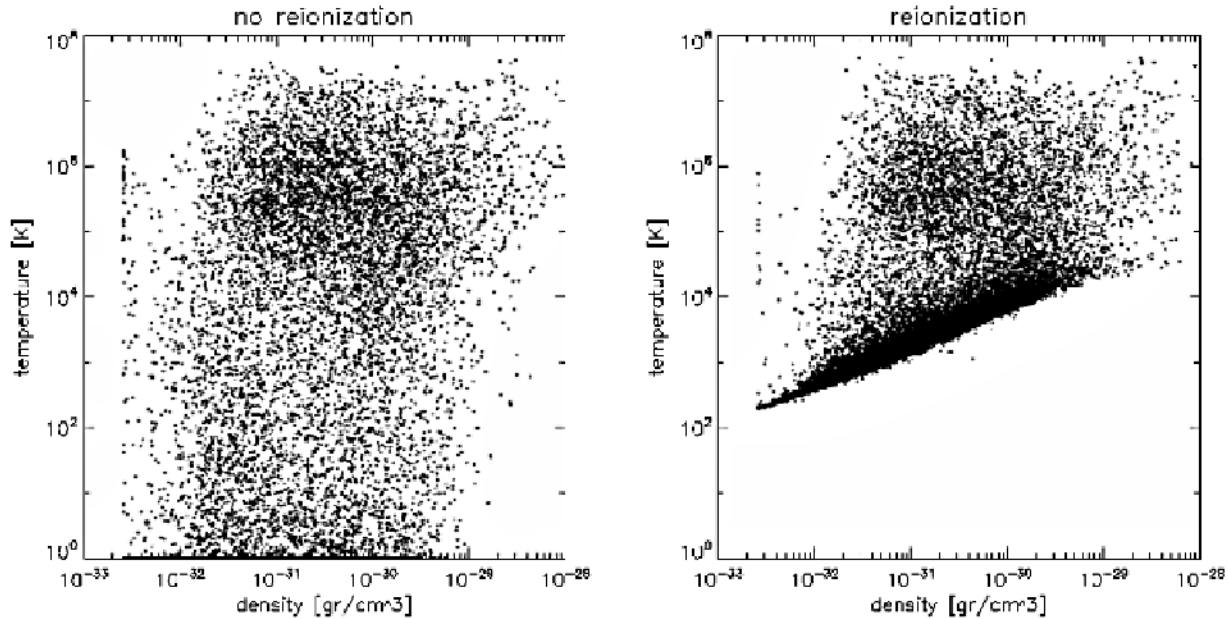}
\caption{Phase diagrams for a box of $80Mpc$, 
from the {\it AD125} run ({\it left}) and
from the {\it CO125} re-simulation ({\it right}).}
\label{fig:phase_col}
\end{figure*}

\section{Cosmological Simulations and Tests.}

\label{sec:sim}

\subsection{General Properties}

In our simulations we have assumed a "Concordance" model, with
density parameters $\Omega_0 = 1.0$, $\Omega_{BM} = 0.044$, $\Omega_{DM} =
0.226$, $\Omega_{\Lambda} = 0.73$, Hubble parameter $h = 0.71$, 
a power spectrum
produced according to the Eisenstein \& Hu (1999) fitting formulas with 
a primordial spectrum normalization $\sigma_{8} = 0.94$, and an initial
redshift of $z=50$.
In order to have a large cluster statistics 
we simulated  a total
volume equivalent to $(145Mpc)^{3} \approx (103Mpc/h)^{3}$ at the fixed 
numerical resolution of 125 kpc.
This total volume was obtained by combining together six (independent)
simulated boxes of 80 Mpc per side. 

A list of all simulations used in our study with their main
properties is listed in Tab.1.
The goal of this study is to investigate  
cosmological shocks with the most simple physical and numerical treatments.
Cosmological shock waves are supposed to be mainly driven by the
assembly of cosmic structures, and therefore 
gravity should be the principal ingredient to model.
Therefore we made massive use of "adiabatic" simulations at 
various resolutions ({\it AD125, AD250,
AD250, AD800}): 
these simulations contain only ''adiabatic'' physics, i.e. they do not
have radiative cooling, UV photo-ionization at early 
epochs, thermal conduction and magnetic fields. 
These simulations are the starting point to investigate the 
effects on the properties of shocks
driven by the adoption of a more
complex physical modeling.
In particular, the re-ionization process
has the important effect of increasing the temperature
(and the sound speed) of cosmic baryons in the low temperature regions,
and thus this is the first additional ingredient to take into account. 
Therefore we re-simulated at two spatial resolutions one of our
six 80 Mpc boxes with the Haardt \& Madau (1996) re-ionization model 
plus radiative cooling ({\it CO125, CO250}) and 
used the outputs of these simulations to extract a recipe 
to mimic the effect of re-ionization in post-processing in adiabatic 
simulations.
Finally, we perform simulations with
a different $\sigma_{8}$ parameter, in order to study how
this parameter can affect
our results ({\it S8250} simulation).

We discuss the effect of re-ionization
and cooling in Sections\ref{sec:reionization} \& \ref{subsec:reion},
and the effect of the numerical resolution
and of $\sigma_{8}$ in Appendix (Sects.A and B).

\subsection{Properties of the Simulated Galaxy Clusters}
\label{subsec:gc}

The aim of this Section is to present the sample of galaxy clusters obtained 
from our simulations and to briefly discuss their main properties.

A cluster reconstruction procedure, based on total over-density criteria
(e.g. Gheller, Moscardini \& Pantano 1998), 
has been applied to the outputs of the
various simulations at different cosmological times, providing a population of
synthetic galaxy clusters which can be followed during time. The overall 
{\it AD125} simulation at $z=0$ consists of 113 galaxy 
clusters with total virial masses 
$\geq 3\cdot 10^{13}M_{\odot}/h$. 

The cumulative mass function of all the halos in our sample is reported in 
Fig.\ref{fig:massfunc} 
and shows an overall good agreement with the Sheth \& Tormen (1999) 
mass function for $M<3\cdot 10^{14}M_{\odot}/h$.
The relevant deficit of halos with $M \geq 3\cdot 10^{14}M_{\odot}/h$
is likely due to the artificial cut-off in the over-density power 
spectrum at long wavelengths in our 80 Mpc simulations
(Bagla \& Ray 2005; Bagla \& Prasad 2006). 
Massive galaxy clusters are expected to be the most important regions
in which kinetic energy is dissipated by shocks (in thermalisation 
and CR acceleration). 
Because a mass deficit in cluster halos  may introduce a deficit in the energy 
processed by shocks in the total simulated volume, this should be taken 
into account when we compare our results with previous studies 
(e.g. Ryu et al. 2003; Pfrommer et al.2006; Kang et al.2007). 

In Fig.\ref{fig:M_T} we report 
the scaling between the total mass and the emission--weighted 
temperature within $r_{500}$ derived for our population of clusters 
{\footnote {$r_{500}$ is the radius encompassing a mean total over-density 
of 500 times the cosmic mean density, and roughly corresponds to 0.5 the
virial radius of galaxy clusters in a $\Lambda$CDM cosmology.}}.
Points are slightly above the self similar scaling found by Borgani 
et al.(2004) although they are consistent within a $\sim 15$ per 
cent scatter; this is true also for the additional
15 halos in the {\it CO125} data set (with radiative cooling 
and re--ionization). 

Overall Figs.\ref{fig:massfunc} and \ref{fig:M_T} show 
that the basic statistical and physical 
properties of our galaxy clusters are in line with those from other 
cosmological numerical simulations.

\begin{figure}
\includegraphics[width=0.48\textwidth]{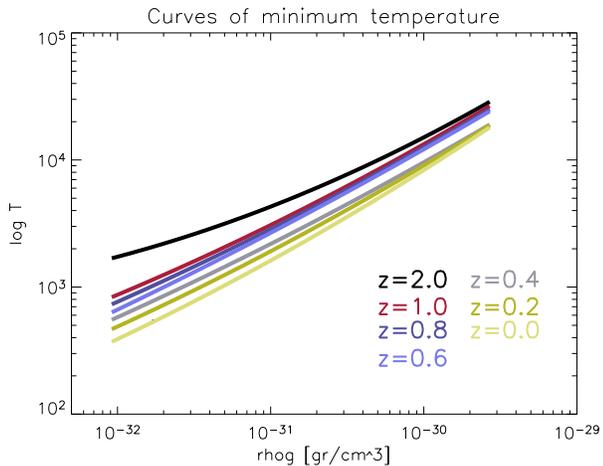}
\caption{Time evolution of the best--fitting relation for the minimum 
temperature of the {\it CO125} run,
from $z=2$ down to $z=0$. Best fit parameters for these curves are reported 
in Tab.2.}
\label{fig:min_t}
\end{figure}

\begin{table}
\label{tab:fit}
\caption{Fit parameters for the minimum temperature for the
outputs of the {\it CO125} run.} 
\centering \tabcolsep 4pt 
\begin{tabular}{c|c|c|c}
 Redshift & $Log T_{0}$ & c1 & c2  \\ \hline
 $z=2.0$ & $3.2383 \pm 0.0032$ & $0.3198 \pm 0.0061$ & $0.0749  \pm 0.0025$ \\
 $z=1.0$ & $2.9388 \pm 0.0010$ & $0.5056 \pm 0.0020$ & $0.0335  \pm 0.0008$ \\
 $z=0.8$ & $2.8846 \pm 0.0009$ & $0.5358 \pm 0.0016$ & $0.0361  \pm 0.0007$ \\
 $z=0.6$ & $2.8288 \pm 0.0011$ & $0.5773 \pm 0.0022$ & $0.0267  \pm 0.0008$ \\
 $z=0.4$ & $2.7628 \pm 0.0019$ & $0.5437 \pm 0.0038$ & $0.0330  \pm 0.0015$ \\
 $z=0.2$ & $2.6889 \pm 0.0016$ & $0.5517 \pm 0.0033$ & $0.0409  \pm 0.0013$ \\
 $z=0.0$ & $2.5904 \pm 0.0016$ & $0.5711 \pm 0.0032$ & $0.0469 \pm 0.0001$ \\
\end{tabular}
\end{table}

\begin{figure*}
\label{fig:distrib_ev}
\includegraphics[width=0.45\textwidth]{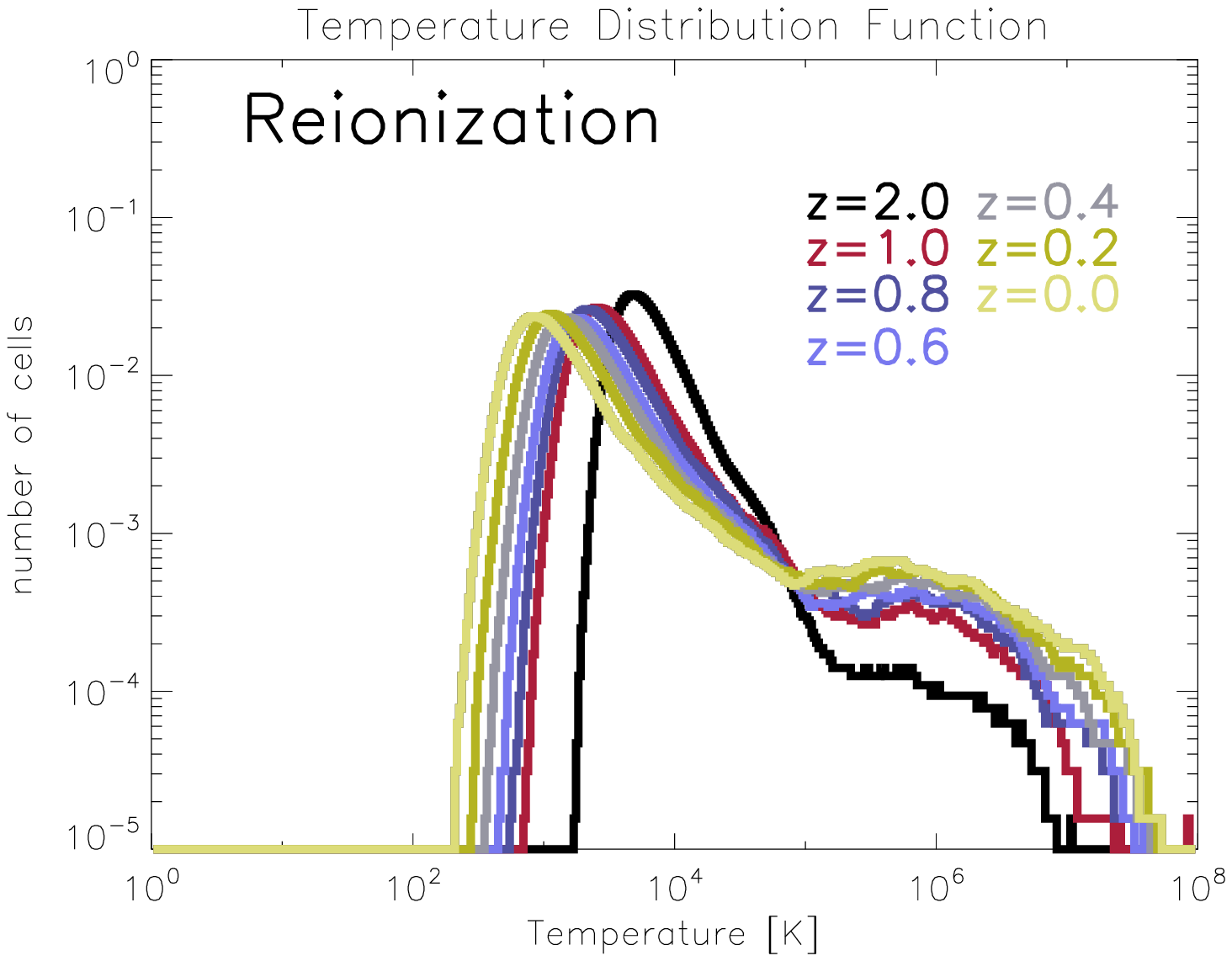}
\includegraphics[width=0.45\textwidth]{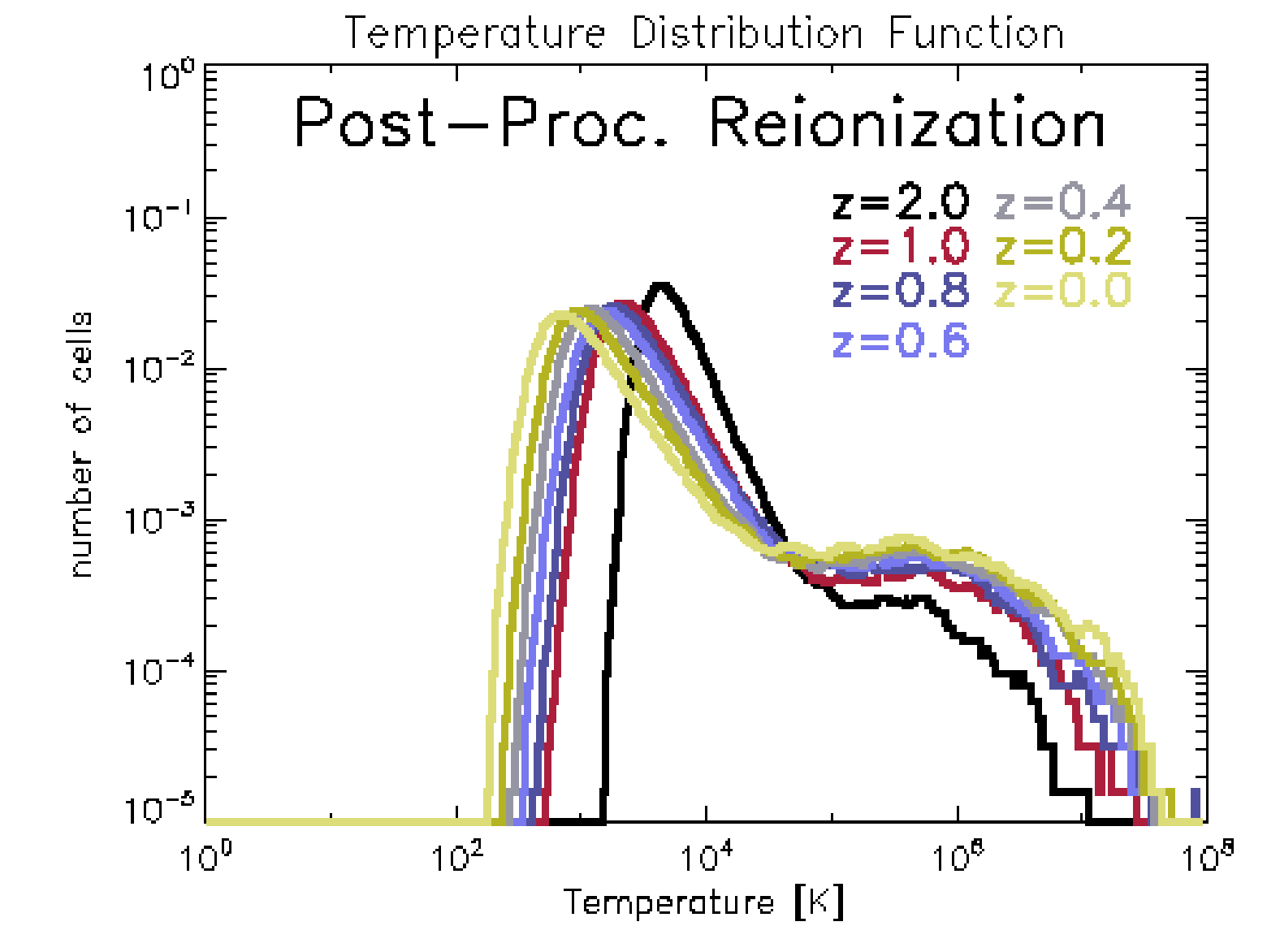}
\caption{Time evolution of the temperature distribution from
$z=2$ to $z=0$ (see panels for  color-coding), for the {\it CO125} simulation ({\it Left}) and for the 
corresponding adiabatic simulation {\it AD125} with the  
post--processing treatment
of re-ionization ({\it Right}).}
\label{fig:madau-vs-post}
\end{figure*}

\section{Phase Diagrams and re--ionizing background.}
\label{sec:reionization}

Shocks and compressions driven by the gravitational force 
are the only sources that increase the thermal energy of cosmic baryons in
our adiabatic simulations. 
Baryons far away from collapsing regions
have the lowest temperature that can be potentially affected by the process 
of re-ionization which occurred between $z\sim 30$ and $z\sim 6$ (e.g. Fan, Carilli \& Keating 2006).
This process heats up the medium, increasing the speed of sound in the
lowest temperature regions and this affects our 
estimate of the Mach number of shocks. Therefore a 
study of cosmological shock waves
must deal with the influence of a re--ionizing background
(Haiman \& Holder 2003; Loeb \& Barkana 2005; Mellema et al.2006).

The re-ionization scheme available in ENZO is 
linked with a treatment of radiative cooling, which is 
computed by assuming an optically thin gas of primordial 
composition, in collisional ionization equilibrium, 
following Katz, Weinberg \& Hernquist 
(1996). The time--dependent UV background is  
introduced according to Haardt \& Madau (1996) and it 
models the effect of a population of quasars that re-ionizes 
the universe at $z\approx 6$. 
The implementation of run--time re-ionization is more 
expensive in terms of memory usage compared to non--radiative
simulations, and we thus 
applied it only in two re--simulated data--sets
({\it CO125, CO250}).
The effect of a re--ionizing background 
can also be modeled in the post--processing phase by 
increasing the temperature of each cell in the simulation.
This has been done in Ryu et al.(2003), where a constant value of $T=10^{4}K$ 
was imposed to the simulated volume
at $z=0$.
This may correctly reflect the complete re--ionization 
inside halos at present epoch (Haiman
\& Holder 2003), however it may overestimate the mean temperature 
of baryons far away of the most massive cosmic structures.
Figure \ref{fig:phase_col} 
shows the phase diagram in one $(80Mpc)^{3}$ simulation from the
{\it AD125} data set, and in its re-simulation with cooling and 
re-ionization, {\it CO125}. 
Re--ionization efficiently removes the coldest phase of the baryons 
and a forbidden region in the
$\log T-\log \rho$ space forms (where $T$ and $\rho$ are gas temperature 
and density, respectively), which actually traces the lower bounds
of the Warm Hot Intergalactic Medium (Katz, Weinberg \& Hernquist
 1996; Cen \& Ostriker 1999; Valageas, Schaeffer \& Silk 2002; 
Regan, Haehnelt \& Viel 2007). 
This lower bound is evolving with time, as shown if Fig.\ref{fig:min_t}
where we report the fit to the value 
of the 15 per cent percentile in the distribution of temperature in the cells 
for different density bins. We also checked that 
variations in the percentile (up to $\sim 50$ per cent) does not 
significantly affect the results.
By restricting to baryon densities in the range
$10^{-32} \leq \rho \leq 3 \cdot 10^{-30} gr/cm^{3}$
we obtain best fits with a second order polynomial :

\begin{equation}
\log ( {{T_{min}}\over{ T_{0}}})=
c_{1} \log ({{\rho}\over{\rho_{0}}})+
c_{2}(\log ({{\rho}\over{\rho_{0}}}))^{2},
\label{eq:fit2}
\end{equation}

where $\rho_{0}=10^{-32}$gr cm$^{-3}$.
The best fit parameters for each redshift are 
reported in Table 2.
At moderate redshift ($z\leq 1$) all curves can
also be approximated with a simple power law, $T_{min} \propto \rho^{0.6}$
(consistent with Valageas et al.2002), and with 
a normalization decreasing with time. 
In particular, at $z=0$ the minimum temperature of baryons is
given by:

\begin{equation}
T_{min}(K)= 450 \,\,( {{\rho}\over{\rho_{0}}})^{0.60},
\label{eq:ion}
\end{equation}

which is indeed consistent with Eq.21 in Valageas et al. (2002). 
By using the {\it CO250} data set we also verified that these fits 
do not change with spatial resolution.

We use the fitting formulas in Tab.2 
to increase the temperature of baryons in our adiabatic simulations 
in the post-processing analysis. 
In Fig.\ref{fig:madau-vs-post}
we show the evolution with time of the temperature distribution
function, in case of adiabatic sub--sample of the {\it AD125} simulation with
post--processing treatment 
and in the case of {\it CO125} simulation
with run--time re-ionization. 
The agreement between the two distributions demonstrates
the validity of our post--processing approach that will 
be applied in the following to the full set of
our adiabatic simulations ({\it AD125, AD250, AD500} and {\it AD800}).

\begin{figure}
\includegraphics[width=0.45\textwidth]{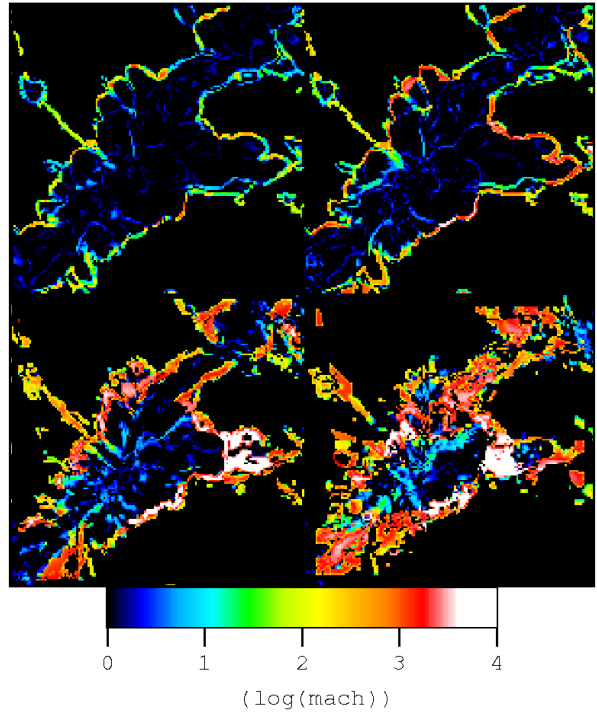}
\caption{Maps of reconstructed Mach numbers using the VJ method based on jumps
of 1 cells (top left), 2 cells (top right), 4 cells (bottom left) 
and 8 cells (bottom right). The image is $20Mpc$ per side, 
the width along the line of sight is $125kpc$.}
\label{fig:lags}
\end{figure}

\begin{figure}
\includegraphics[width=0.48\textwidth]{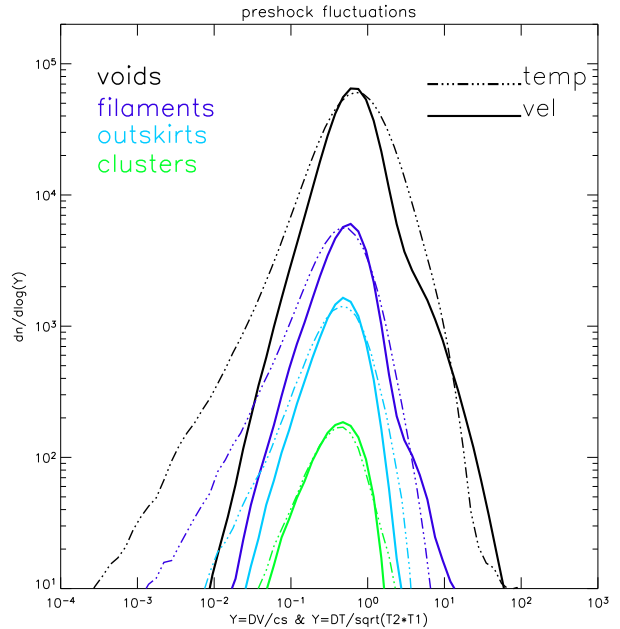}
\caption{Distribution of temperature fluctuations 
({\it dot--dash lines}) and  velocity fluctuations 
({\it solid lines}) for non-shocked cells in the simulation, at four different 
over-density regimes (see labels within the panel). Data are taken from one box of side $80Mpc$ of our $AD125$ run.}
\label{fig:DeltaT_histo}
\end{figure}

\section{Shock--Detecting Methods}

\label{sec:algo}

\subsection{Basic Relations}

The passage of a shock in a simulated volume leaves its imprint as 
a jump in all the thermodynamical variables. Under the simple assumption that
the pre--shocked medium is at rest and in thermal and pressure equilibrium, 
the pre--shock and post--shock values for any of the hydrodynamical variables
(density, temperature and entropy) is uniquely related
to the Mach number, $M=v_{s}/c_{s}$, $v_{s}$ being the shock speed 
in the region and $c_{s}$ the sound speed ahead of the shock itself. 
The Rankine--Hugoniot jump conditions contain
all the information needed to evaluate $M$; if the adiabatic
index is set to $\gamma = 5/3$ one has the well known relations
(e.g. Landau \& Lifshitz 1966):

\begin{equation}
\frac{\rho_{2}}{\rho_{1}}=\frac{4M^{2}}{M^{2}+3},  
\label{eq:dens}
\end{equation}

\begin{equation}
\frac{T_{2}}{T_{1}}=\frac{(5M^{2}-1)(M^{2}+3)}{16M^2}
\label{eq:temp}
\end{equation}

and

\begin{equation}
\frac{S_{2}}{S_{1}}=\frac{(5M^{2}-1)(M^{2}+3)}{16M^{2}}
(\frac{M^{2}+3}{4M^{2}})^{2/3},  
\label{eq:entropy}
\end{equation}

with indices $1,2$ referring to pre and post--shock quantities,
respectively, and where the entropy $S$ is $S=T/\rho^{2/3}$.  

The Mach number can be obtained from the jumps in one of the
hydro dynamical variables (Eqs.\ref{eq:dens}--\ref{eq:entropy}) 
or from a combination of them.
It is well known that the value of the density jump saturates at 
$\rho_2/\rho_1 = 4$ in the case of relatively large Mach numbers 
(Eq.\ref{eq:dens}) and thus strong shocks cannot be 
constrained from density jumps. 
For this reason temperature and entropy jumps provide the most 
effective tools to measure $M$.

Eqs.\ref{eq:dens}--\ref{eq:entropy} 
describe shock discontinuity
in the case of an ideal shock. 
In practice measuring the Mach number of the shocks in simulations 
is more problematic. 
Matter falling in the potential wells
may have chaotic motions and the temperature distribution 
is usually patchy due to the continuous accretion and mixing
of cold clumps and filaments into hot halos. 
All these complex behaviors establish velocity, temperature
and density discontinuities across the cells in the simulated box.
In a post-processing analysis this is expected to modify irreparably 
the strength of the jumps in the thermodynamical variables in the
shocked cells with respect to that 
expected in the ideal case (Eqs.\ref{eq:dens}--\ref{eq:entropy}). 
Consequently the estimate of the Mach
number from these equations is subject to unavoidable uncertainties (see
Sec.6.2).

\subsection{The Temperature Jumps Method}
\label{subsec:Ryu0}

The analysis of jumps in temperature is commonly adopted to measure
the strength of shocks in Eulerian cosmological simulations 
(e.g., Miniati et al. 2001; Ryu et al. 2003).

Cells hosting a possible shock pattern are preliminary tagged by means 
of two conditions: 

\begin{itemize}
\item $\nabla T \cdot \nabla S > 0$;
\item $\nabla \cdot {\rm v} < 0$;
\end{itemize}

An additional condition on the strength of the temperature 
gradient across cells is also customary requested, e.g.

\begin{itemize}
\item $\mid \triangle log T \mid \geq 0.11$;
\end{itemize}

\noindent
(specifically $\mid \triangle log T \mid \geq 0.11$ 
filters out shocks with a Mach number $M < 1.3$, Ryu et al.2003); however 
in the following
we neglect this third condition, in order to have a better
comparison with the results obtained with the VJ method (see below).

The shock discontinuity in the simulation is typically spread over a few cells, thus
following Hallman et al.(2004) for each  patch of candidate 
shocked cells we define the shock center with the position of the
cell in the shocked region where $\nabla \cdot v$ is minimum and
calculate the Mach
number of the shock from Eq.\ref{eq:temp}, where $T_2$ and $T_1$
are the post and pre--shock temperature across the shock region.
The Mach numbers measured along the three coordinate axes are finally
combined to compute the Mach number of the shocked cell:
$M = (M_{x}^{2}+M_{y}^{2}+M_{z}^{2})^{1/2}$. 

More specifically, in the case the shock-jump is assumed to happen
in 1 cell, $T_2$ is the temperature of the shock center, while
in the case that the jump is spread over 2, 4, 6, {\ldots} 2\,{\it n} cells
$T_2$ and $T_1$ are the temperatures of the two cells at distance of {\it n}
cells
(in opposite direction) from the center of the shock.

In the following we will refer to
this method as the {\it TJ} method.

\begin{figure*}
\includegraphics[width=0.95\textwidth]{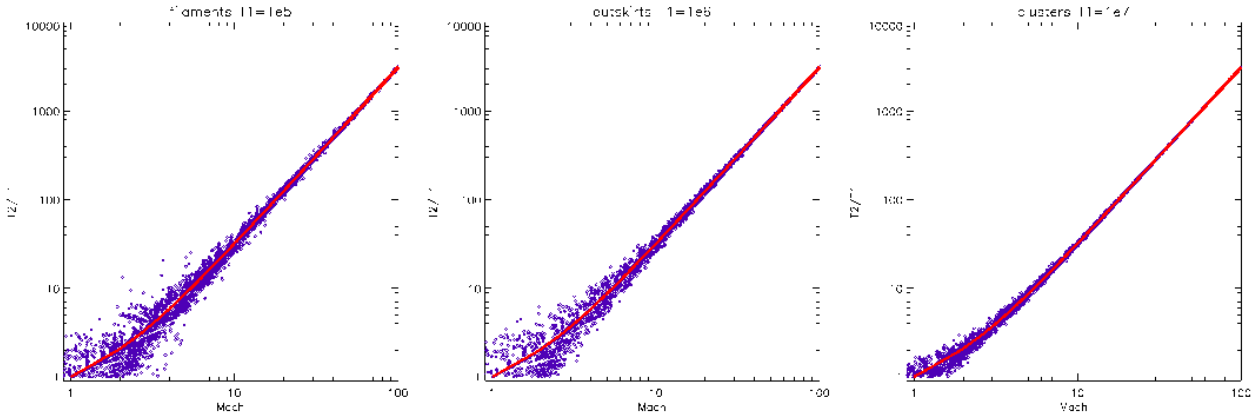}
\caption{Scatter plots for the value of Mach numbers from Monte Carlo
extracted populations of non shocked cells, in filaments, outskirts and clusters (see text).
Temperature fluctuations are extracted from the corresponding
distributions in Fig.\ref{fig:DeltaT_histo}. The red curves give the 
exact solution from Eq.\ref{eq:temp}.}
\label{fig:error_TJ}
\end{figure*}

\begin{figure*}
\includegraphics[width=0.95\textwidth]{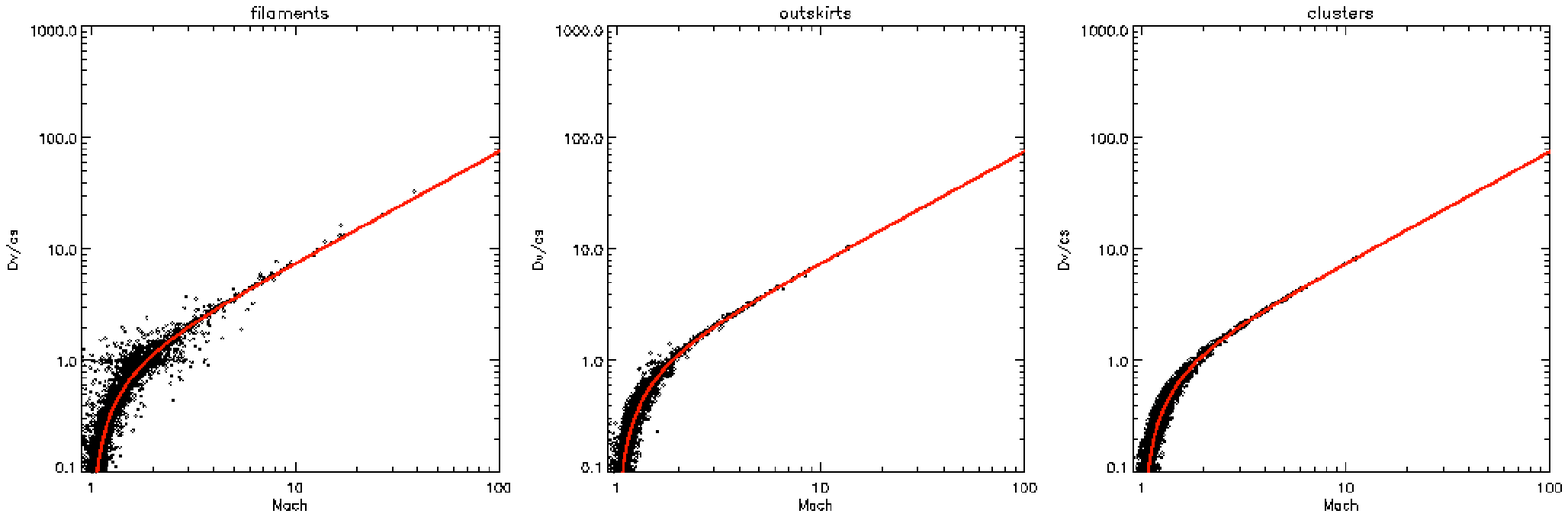}
\caption{Scatter plots for the value of Mach numbers from Monte Carlo
extracted populations of non shocked cells, in filaments, outskirts and clusters (see text.)
Velocity fluctuations are extracted from the corresponding 
distributions in Fig.\ref{fig:DeltaT_histo}. 
The red curves give the solution from Eq.\ref{eq:mach_v}.}
\label{fig:error_VJ}
\end{figure*}

\begin{figure*}
\includegraphics[width=0.95\textwidth]{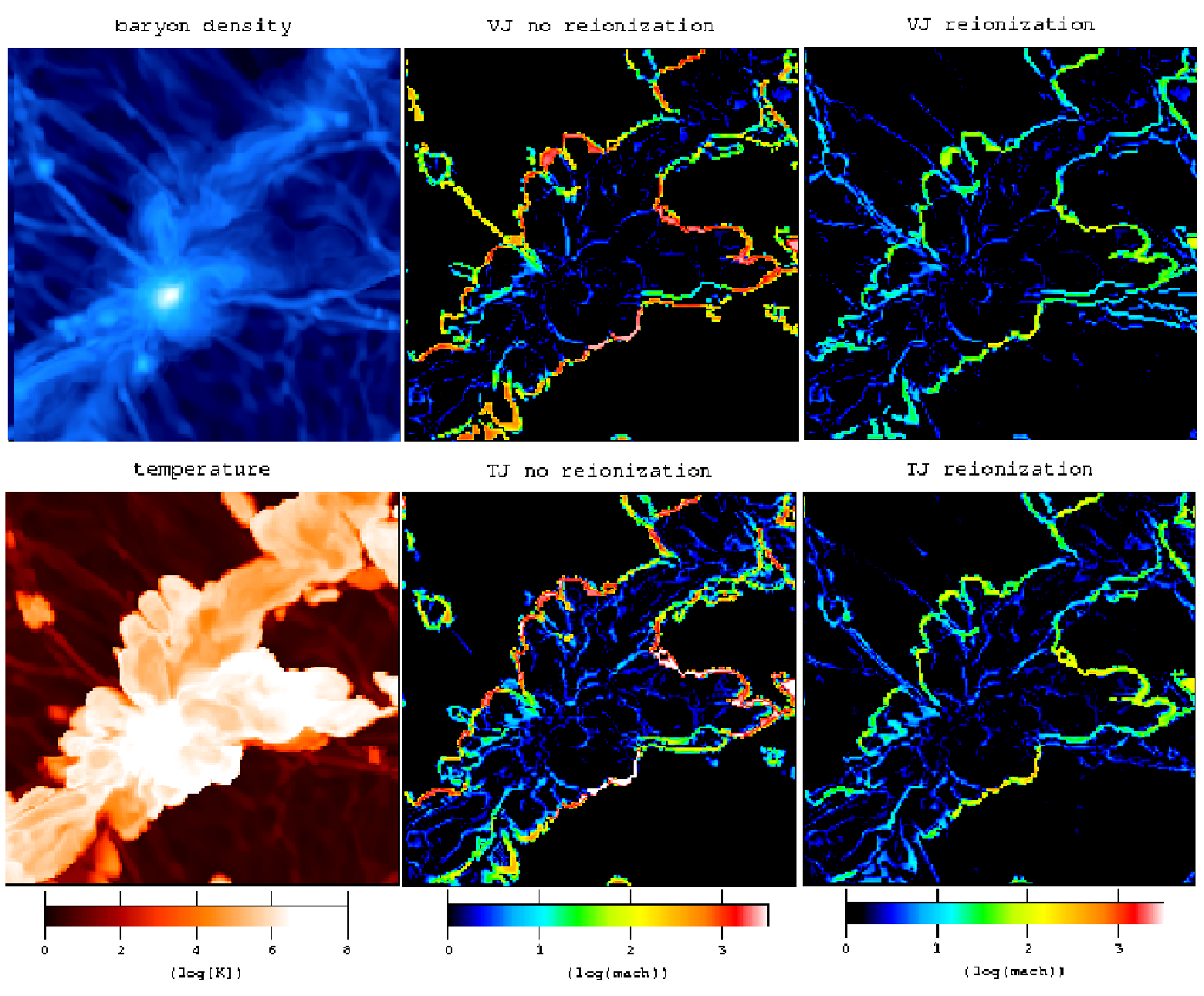}
\caption{Maps of a $20Mpc$ region centered on a $M_{tot}\sim 10^{14}M_{\odot}$ galaxy
cluster, with $125kpc$ of cell resolution (the line
of sight width is one cell). Left column shows the baryon density (top panel) and baryon
temperature (bottom panel), in this case taken from the same region of the $CO125$ run); the central and the right columns show
the maps of Mach number according to the VJ scheme (bottom panels) and to the
TJ scheme (bottom panels). 
Shocks are shown both in the case of no-reionization ($AD125$, central column) and for 
the case of re-ionization ($CO125$, right column).} 
\label{fig:fix_reion}
\end{figure*}

\subsection{The Velocity Jump Method.}
\label{subsec:vj}

Conservation of momentum in the reference frame of 
the shock yields: 

\begin{equation}
\rho_{1}v_{1}=\rho_{2}v_{2},
\label{eq:momentum}
\end{equation}

with the same notation used in Eqs.\ref{eq:dens}--\ref{eq:entropy}.
In the ideal case in which the pre-shocked medium is 
at rest and in thermal and pressure equilibrium, 
the passage of a shock with velocity $v_s$ leaves a clear imprint 
as a velocity difference, 
$\Delta v$, between the shocked and pre--shocked cells.
The relationship between $\Delta v$ and Mach number in the
case of hydrodynamical shocks 
can be obtained by combining Eq.\ref{eq:momentum} with Eq.\ref{eq:dens} and by 
transforming the velocities from the shock frame 
to the Lab frame :

\begin{equation}
\Delta v =\frac{3}{4}v_{s}\frac{1-M^{2}}{M^{2}}.
\label{eq:mach_v}
\end{equation}

where $v_{s} = M c_{s}$ and $c_{s}$ is the sound velocity computed
in the pre--shocked cell.

The procedure we adopt to identify shocks and
reconstruct their Mach numbers 
is the following :

\begin{itemize}
\item we consider candidate shocked cells those with 
$\nabla \cdot {\rm v} < 0$ (calculated as 3--dimensional velocity 
divergence to avoid confusion
with spurious 1--dimensional compressions that may happen
in very rarefied environments);

\item since shocks in the simulation are typically spread over
a few cells, as in the case of the TJ method, we define the shock center 
with the position of the
cell in the shocked region with the minimum divergence;

\item we scan the three Cartesian axes with a one--dimensional procedure 
measuring the velocity jump, $\Delta v_{x,y,z}$, between a few
cells across the shock center. 
In the case the shock-jump is assumed to happen in 1 cell $\Delta v_{x,y,z}$ 
is calculated between the shock center and the pre-shocked cell, while 
in the case that the jump is spread over 2, 4, 6, {\ldots} 2\,{\it n} cells 
$\Delta v_{x,y,z}$ is calculated between two cells at distance of {\it n} 
cells (in opposite direction) from the center of the shock;

\item the Mach number of the shock is given by 
Eq.\ref{eq:mach_v}, where the sound speed is that of the
pre-shock region (the cell with the minimum temperature);

\item we finally assign to shocked cells a Mach number 
$M = (M_{x}^{2}+M_{y}^{2}+M_{z}^{2})^{1/2}$, that minimizes 
projection effects in the case of diagonal shocks, and restrict
to shocks with $M>1$.

\end{itemize}

\noindent
In the following we refer to this procedure as 
the velocity jump (VJ) method.

\section{{\bf Uncertainties in Shock Detecting Schemes}}
\label{sec:comparison}

In this Section we discuss the uncertainties of the methods
presented in the previous Section, discuss the effect of the
re-ionization on the characterization
of cosmological shocks and briefly compare
results from the VJ and TJ approaches.

\subsection{Reconstruction of the shock discontinuity}

Although the shock discontinuity in ENZO is found to be well reconstructed
within 2-4 cells (e.g., Tasker et al. 2008), the risk that comes
from the application of procedures based on cell-to-cell jumps (or
jumps between a few cells) is to underestimate the
value of the Mach number of the shock.
We performed several shock--tube tests with ENZO with the
same numerical setup used in the cosmological simulations, 
in order to evaluate the
convergence of the value of the shock Mach number with the number
of cells used to calculate jumps.
We find that a reasonable convergence, within 10-30\% for $M< 10$,
is already obtained with the VJ method in the case that the velocity jump
is evaluated across three cells ($n=1$, where $n$ is the distance, in term
of cells,
between the shock center and the pre or post-shock cells, Sects.~5.2-5.3), 
and that convergence is reached
for $n \leq 2$.
On the other hand, the velocity pattern in cosmological simulations
is complex and the risk of procedures based
on jumps evaluated with large $n$ in our simulations 
is to mix together signals produced by different shocks, 
and also to be affected by gradients in thermodynamical variables
due to clumps of baryonic matter produced by the process of
structure formation.
In Fig,~\ref{fig:lags} we show a map of the Mach numbers obtained with the VJ
method for a galaxy cluster in the $AD125$ run, by assuming a cell-to-cell (two cells) velocity jump,
and $n=1$, $n=2$ and $n=4$ jumps.
It is clear that for $n \geq 2$ (jumps based on $\geq$5 cells,
$\geq$625 kpc)
different shock-patterns and clumps of gas matter start to be mixed together
 and shocks become
poorly characterized.

Similar results are found in the case of the TJ method, thus 
we conclude that reconstructing the shock discontinuities in our
numerical simulations with $n=1$ (jumps based on 3 cells) provides
the best compromise.

\subsection{Uncertainties in the TJ and VJ methods}

As already pointed out in Sect.\ref{sec:algo}, a major limitation 
of the analysis of shocks in post processing 
comes from the fact that the dynamics and thermodynamics of the gas in
the simulations is more complex than in the ideal case in which 
Eqs.\ref{eq:dens}--\ref{eq:entropy} and \ref{eq:mach_v}
are derived.
In this sub-section we discuss the uncertainties that come out  
without including the modeling of re-ionization in our procedure.

\subsubsection{TJ method}
\label{subsec:tjun}

The temperature distribution in simulations is very complex 
and temperature gradients across non--shocked cells 
are common by--products of the structure formation process.
The passage of a
shock in a medium with a complex temperature distribution
partially smooths out pre-existing gradients in the thermodynamical
variables and creates new shock--induced discontinuities.

One possibility to estimate
the level of uncertainties in the application of the TJ method
is to introduce a passive modification of
the post--shock temperature
in Eq.\ref{eq:temp}
according to the value of a typical cell
to cell temperature jumps across non shocked cells, and
compare the
resulting Mach number with that from Eq.\ref{eq:temp} in its original form.
Obviously this procedure assumes that these jumps are representative
of pre-existing temperature gradients, where
shocks are presently found, still there is no clear argument
for which this unavoidable assumption is not statistically reasonable.

\noindent
We consider as non--shocked
cells those that do not satisfy, at the same time,
the TJ and VJ criteria for shocked cells, and
extract the values of their cell to cell temperature
jumps, $\delta T$, in different cosmic
environments from the {\it AD125} simulation at $z=0$.
To follow a very conservative procedure we consider only
temperature jumps across a sub--sample of non shocked cells 
that are at least three 
cells far away from any shocked cell.

We characterize the cosmic environment
by means of the total matter density in cells :

\begin{itemize}

\item {\bf $0.01 \leq \rho_{tot}/\rho_{cr}<3$}: voids and under-dense 
regions,

\item {\bf $3\leq\rho_{tot}/\rho_{cr}<30$} : filaments and sheets,

\item {\bf $30\leq\rho_{tot}/\rho_{cr}<50$}:  cluster outskirts,

\item  {\bf $\rho_{tot}/\rho_{cr}\geq 50$}: galaxy clusters.

\end{itemize}

where $\rho_{tot}=\rho + \rho_{dm}$ is the total matter
density and $\rho_{cr}$ is the critical density of the universe. These
are expected to mark the different kind of structures
of the cosmic web (e.g. Dolag et al.2006; Shen et al.2006).

\noindent
In Fig.\ref{fig:DeltaT_histo}
we plot the differential distribution of (the module of)
temperature jumps across the considered sub-sample of
non--shocked cells, $\delta T$,
normalized to a reference value of the local temperature,
$\sqrt{T_{2}\cdot T_{1}}$, for the different density regimes.
The peak of the distribution is found at 
$\delta T/\sqrt{T_{2}\cdot T_{1}} \approx 0.5$.
In the case of filaments and voids the distributions extend
at larger values, although large temperature scatters,
$\delta T/\sqrt{T_{2}\cdot T_{1}} \approx 10$, are only found in voids, where they represent less then 1 percent of all cells{\footnote{We also note that rare large temperature jumps in low density environments, not associated with shocks, might come from numerical artifacts in regions where gas flows are supersonic and the computation of the internal energy is more subject to numerical uncertainties.}}.

 extremely rare 
in the case of filaments and are found for only a few percent of 
the cells in the voids.

\noindent
For the values of $T_{1}$ representative of clusters,
outskirts and  filaments in our simulation,
we allow $T_{2}$ to vary and run Monte Carlo
extractions of $\delta T$ extracted across non--shocked cells
with temperature $T_{2}$ in the same environment.
We then compared the
estimate of the shock Mach number by Eq.\ref{eq:temp}
with the one obtained by using:

\begin{equation}
\frac{T_{2}\pm | \delta T |}{T_{1}}=\frac{(5M^{2}-1)(M^{2}+3)}{16M^2}.
\end{equation}

Figure 7 shows the typical scatter introduced in the
$T_{2}/T_{1}$ vs $M$ plane by the presence of realistic (i.e. measured
in non--shocked cells in our simulations) pre--shock fluctuations in
the temperature, for different cosmic environments.
The red line shows the ideal case: given a ratio $T_{2}/T_{1}$
the degree of uncertainty on $M$ due to the presence of
pre--shock fluctuations in the simulation
can be grossly evaluated by an
horizontal cut across the distribution of the data points.
This scatter increases as the Mach number decreases and,
at a given Mach number, it is typically smaller in 
environments with larger over-density.

\begin{figure}
\includegraphics[width=0.48\textwidth]{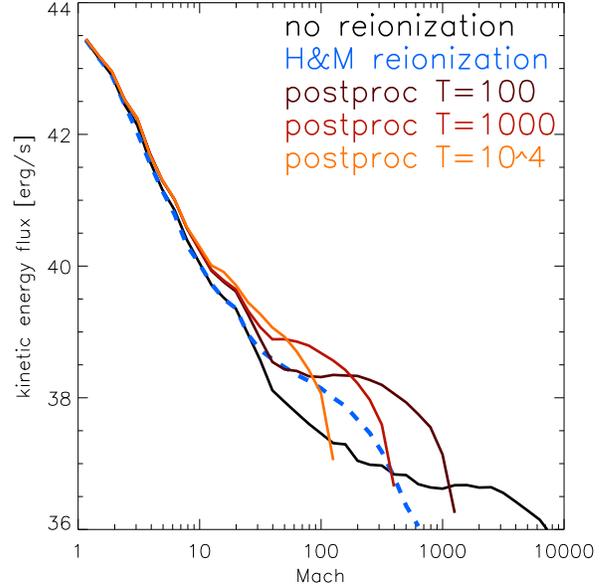}
\caption{Distribution
of kinetic energy flux in shocks according to the VJ method, 
for a cubic volume of side $40Mpc$ and resolution $125kpc$. 
Curves are drawn for the case without
re-ionization ({\it black solid}), for the Haardt \& Madau (1999) 
re-ionization scheme applied in post-processing ({\it blue dashed}) 
and for different choices 
of a fixed $T_{floor}$ temperature
floor (color coding is labeled in the panel).}
\label{fig:rhov_cfr}
\end{figure}

\begin{figure*}
\includegraphics[width=0.95\textwidth,height=0.4\textwidth]{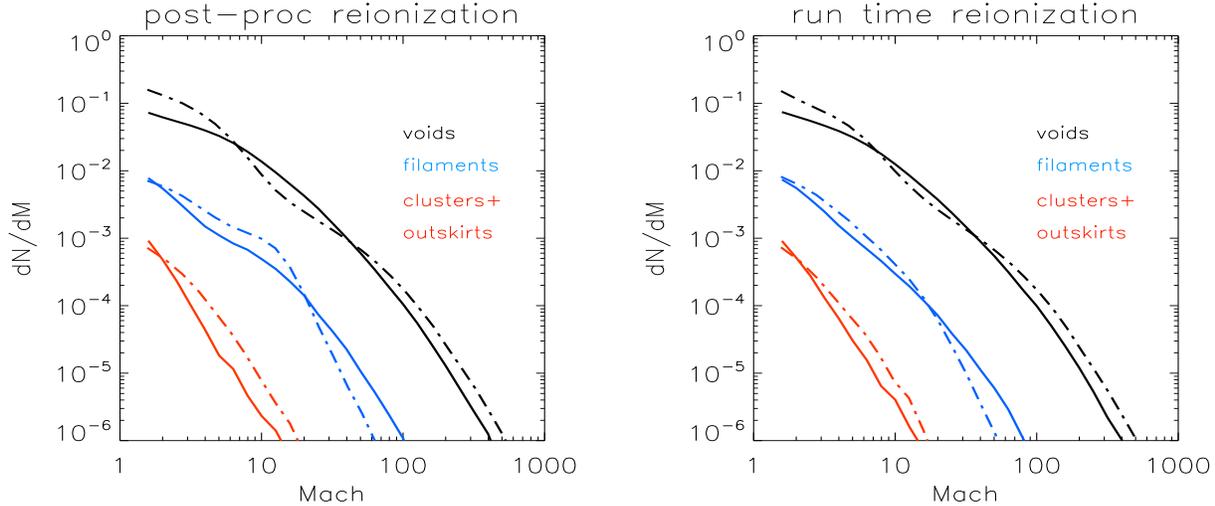}
\caption{Distributions of shock Mach numbers at different cosmic environments, for a  $(80Mpc)^3$ volume of the $AD125$ run with post-processing re-ionization (left) and for the $CO125$ run (right), with the VJ method ({\it solid lines}) and with the TJ method ({\it dot-dashed lines}).} 
\label{fig:histo_vjtj}
\end{figure*}

\subsubsection{VJ method}
\label{subsec:vjun}

\noindent
Complex velocity fields arise naturally during the
formation of virialised structures in simulations
(Bryan \& Norman 1998, Sunyaev; Norman \& Bryan 2003; Dolag et al. 2005;
Vazza et al. 2006; Iapichino \& Niemeyer 2008) that however 
are expected to be smaller than the velocity
jumps driven by the passage of a shock across the same regions.
A more complex situation is that of non virialised structures
where laminar flows may produce relatively strong velocity
gradients across the cells.
An example is given in Fig.~\ref{fig:DeltaT_histo}, where we report the
differential distributions of
the velocity gradients, $\delta v$, normalized to the maximum value of
the sound speed in each pair of cells, obtained for the same sub-sample 
of non-shocked cells considered in the previous sub-Section.
The distributions peak at $\delta v /c_s \approx 0.5$, although 
tails extending towards larger values are found in the 
distributions of voids and filaments.
These tails are mostly due to velocity gradients measured 
across accelerated laminar flows (where the kinetic energy
of the gas may become even larger than the thermal energy)
that move from low to higher density
regions and are present in a small fraction of the volume of 
filaments and voids. 

\noindent
In order to grossly estimate the strength of the uncertainties
in the case of clusters, outskirts and filaments, 
we follow a procedure similar to that in Sect.\ref{subsec:tjun}.
For these different environments
we fixed a value of $\Delta v /c_s$, run a Monte Carlo extraction
of $\delta v /c_s$ from non shocked cells in the simulations
(Fig.\ref{fig:DeltaT_histo}) and for each 
trial calculated the Mach number from :

\begin{equation}
\Delta v \pm | \delta v | =
\frac{3}{4}\cdot \frac{1-M^{2}}{M}
c_s \left( 1 \pm {{ \delta c_s }\over{c_s}} \right).
\label{eq:mach_v_err}
\end{equation}

This equation accounts for both pre-shock gradients in the velocity
and in the sound speed across non shocked cells.
Gradients in $c_{s}$ are driven by gradients in the temperature 
distribution of the cells and are evaluated by a Montecarlo extraction 
of the temperature variations in Fig.~\ref{fig:DeltaT_histo}.

In Fig.\ref{fig:error_VJ} we report $\Delta v/c_s$ vs $M$ 
from our Monte Carlo extraction compared to
the calculations in the {\it ideal case} (Eq.\ref{eq:mach_v}).
This result should be compared with that in Fig.\ref{fig:error_TJ}
obtained for the TJ method and the degree of uncertainty on $M$
can be grossly evaluated by an horizontal cut across the
distribution of the data points.
As expected, in the case of clusters and outskirts 
the scatter in the two cases is quite similar, although
in the case of weak--moderate shocks crossing filaments 
and outskirts the scatter is less pronounced
than that of the TJ method (Fig.\ref{fig:error_TJ}).

\subsection{Modeling the re-ionization.}
\label{subsec:reion}

The role of re-ionization is of primary importance
to study the properties of shocks outside galaxy clusters.
In adiabatic simulations,
regions far away from intense structure formation are very cold
due to the lack of intense shock heating and due to cosmic expansion.
In these environments, any additional source
of heating (such as re-ionizing radiation from AGN and/or
massive stars feedback) causes a dramatic increase of
the local temperature and sound speed.
Thus the temperature distribution across
cells in these regions is strongly affected by the modeling of the
re-ionization in the simulations, and this implies an additional uncertainty
in the characterization of shocks.
This is expected to be particularly relevant in all shock detecting
schemes where temperature plays a role.

Therefore in this Section we highlight 
the main effect of cosmic re-ionization on shocks
Fig.\ref{fig:fix_reion}
shows the maps of the detected shocks
in a $20Mpc$ cubic region extracted from the {\it AD125}
simulation and centered on a $M_{tot} \sim 10^{14}M_{\odot}$
cluster. Results are reported, by calculating shock-jumps across three cells
(n=1, Sect.~5.2, 5.3),
in the case of no re-ionization and of a Haardt \& Madau (1996)
re-ionization scheme.
As expected the Mach number of shocks decreases in simulations with
re-ionization due to the increase of the sound speed produced 
by the re-ionization background.
This effect is stronger in the cold outermost regions,
while the properties of cosmological shocks in galaxy
clusters are not affected by the re-ionization background.
Re-ionization also allows to better describe shocks around
filamentary structures in low density regions 
that are not seen in the case of 
simulations without re-ionization (Fig.\ref{fig:fix_reion}). 
This is because without re-ionization these regions have temperature 
so small that the temperature floor (1 K) adopted (in the outputs) 
by ENZO artificially affects their temperature distribution.

In Fig.\ref{fig:rhov_cfr} we report the kinetic energy flux through
shocked cells as calculated by means of the VJ method.
The kinetic energy flux, $E_{kin}=\rho v_{s}^{3}/2$, is reported for different
numerical modeling of the re-ionization: three different
temperature floors ($10^{4}K$, $10^{3}$ and $10^{2}K$),
Haardt \& Madau (1999) model, and no re-ionization.
We find that a fixed temperature floors, which is customary used
in several papers to mimic the effect of re-ionization (e.g.
Ryu et al. 2003), produces some {\it artificial} piling up or flattening
in the distributions of the energy flux through
shocks at large Mach numbers.
This is because the temperature background, $T_{floor}$, 
changes artificially the speed of the sound in environments with lower
temperature and the Mach number of shocks in these environments 
is affected by $T_{floor}$, decreasing artificially
with increasing $T_{floor}$.
This further supports the requirement of a physically meaningful treatment 
of re-ionization in a post processing procedure.
As already discussed, the post processing fitting procedure
described in Sect.\ref{sec:reionization}
closely resembles the effect of the physically based
Haardt \& Madau (1999) re-ionization scheme.

\begin{figure*}
\includegraphics[width=0.44\textwidth]{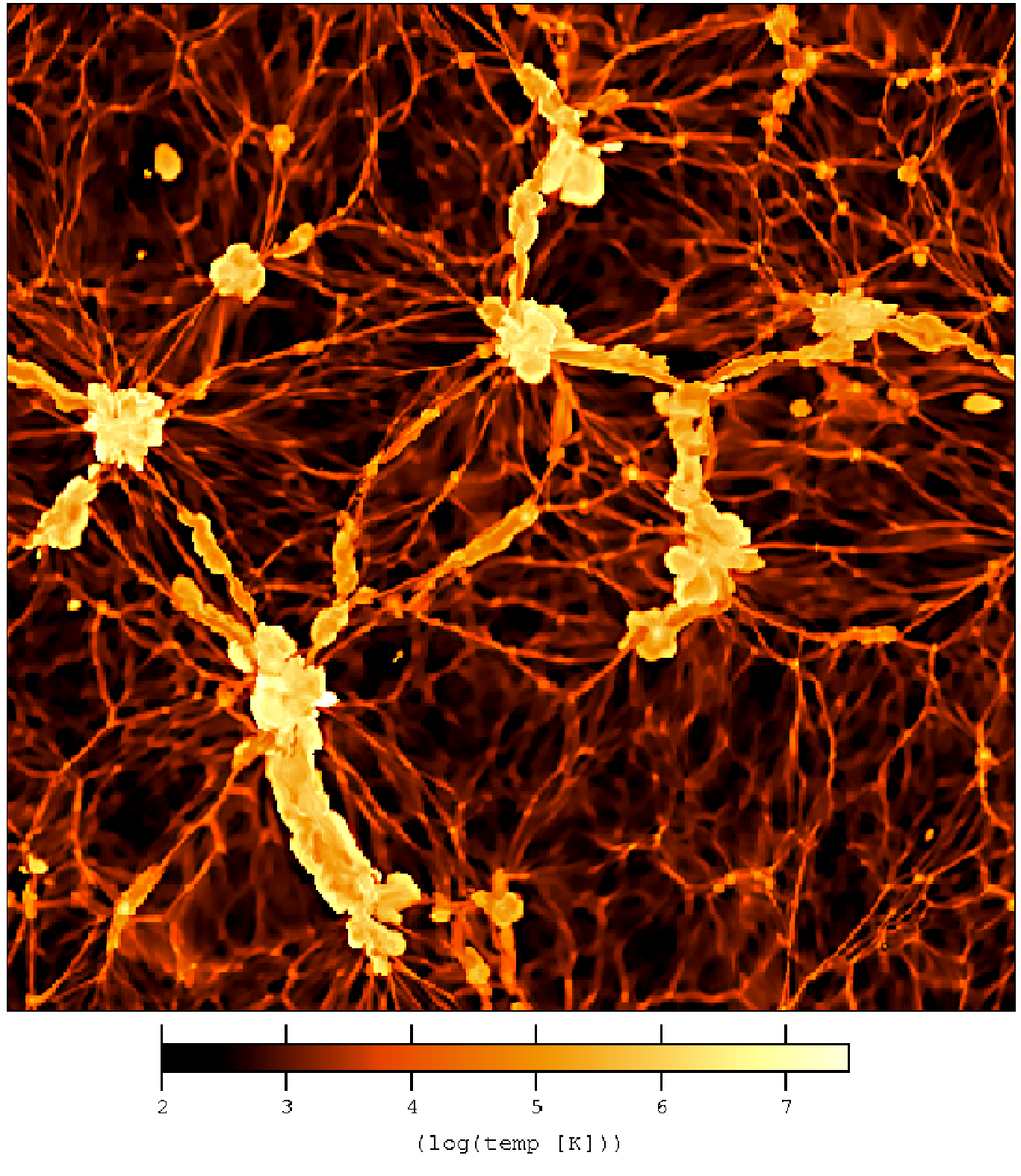}
\includegraphics[width=0.44\textwidth]{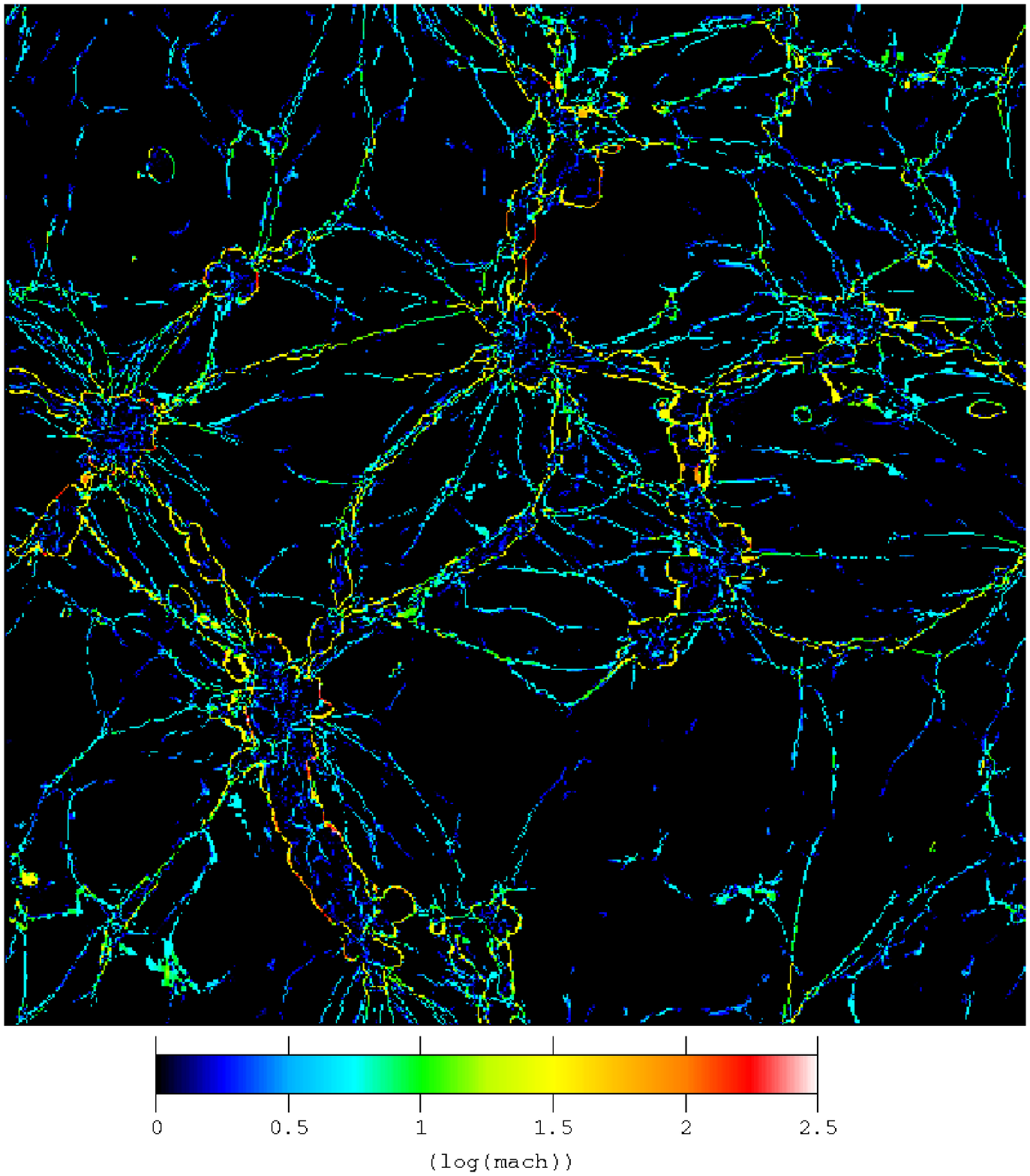}
\caption{Maps of a region of side $80Mpc$ from the
{\it AD125} run with post-processing re-ionization, showing gas 
temperature ({\it left}) 
and Mach number measured with the VJ method  
({\it right}). The width along the line of sight is $125 kpc$.}
\label{fig:mappe}
\end{figure*}

\subsection{Basic Comparison between VJ and TJ methods}
\label{subsec:tjvj}

In this Section we briefly compare the 
results obtained from the VJ and TJ approaches.
Here we focus on results obtained with our
fiducial numerical treatment of the re-ionization as this
treatment will be used in the following of the paper; 
a more detailed comparison is ongoing and will be
presented in a forthcoming paper (Vazza et al. in prep).

In the ideal case the two approaches must select the same population of
shocked cells.
In reality we find that, when shock-jumps are calculated across 3 cells,
about 85 per cent of the shocked cells in our
simulations are selected at the same time by the conditions in 
the VJ and TJ approaches. In the case of clusters and cluster outskirts the
two approaches select the same population of shocked cells, while
these differences typically arise from shocks 
in low temperature regions. 

In the case of clusters and outskirts the velocity and temperature
variations across non shocked cells are relatively small 
(Fig.~\ref{fig:DeltaT_histo}) and this allows constraining the Mach number 
of shocks by means of both the TJ and VJ approach.
Still the statistical
uncertainties for weak shocks with
the TJ method are expected to be 
slightly larger than those with the 
VJ (Figs.\ref{fig:error_TJ} \& \ref{fig:error_VJ}).

A comparison between the statistical description of the
properties of the shocks with VJ and TJ approaches is 
shown in Fig.\ref{fig:histo_vjtj} where we report the Mach number
distribution of shocks extracted from the {\it AD125} 
run with our post-processing treatment of re-ionization and from
the corresponding {\it CO125} with run-time treatment of re-ionization.
Basically the results from the two methods are fairly
similar in the case of clusters and cluster outskirts, and no remarkable
differences are found also in the case of filaments and voids.
This suggests the important point that 
the characterization of shocks in these environments
is statistically solid, as two independent approaches 
lead to basically similar results.
We also note that the Mach number distributions extracted from
simulations with post processing re-ionization are similar to those
produced with run time re-ionization, and this further suggests that
our treatment of re-ionization in post processing allows statistically valid
studies of shocks.

In the next Sections we shall use the VJ method to study shocks properties. 
This is because we believe that
in the case of weak--moderate shocks, especially
in lower density regions, the VJ is less affected by uncertainties
(e.g., Figs.~8-9).

\section{RESULTS}
\label{sec:results}

In this Section we present the main results obtained for
the full set of simulations by making use of the VJ method
and by calculating shock-jumps across three cells, i.e. $n=1$ 
(unless specified).

\begin{figure}
\includegraphics[width=0.47\textwidth]{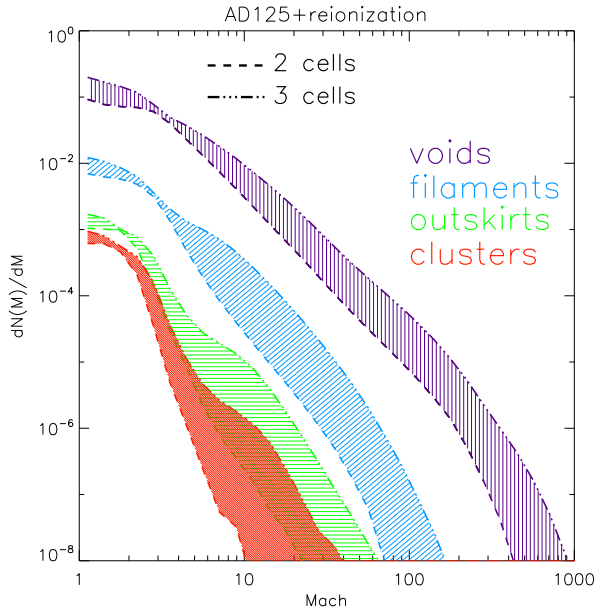}
\caption{Distribution of shocks Mach number for the whole simulated volume of
the $AD125$ with post-processing re-ionization, for different 
cosmic environments. Dot--dashed lines show the distributions
obtained with velocity jumps evaluated across three cells ($n=1$), 
while dashed lines shows distributions obtained with cell--to--cell
velocity jumps ($n=0$).}
\label{fig:histo}
\end{figure}

\begin{figure}
\includegraphics[width=0.48\textwidth]{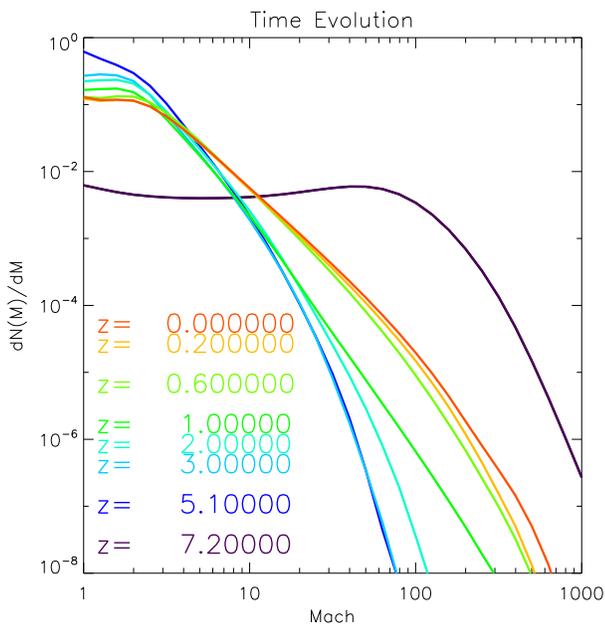}
\caption{Time evolution of the Mach number distribution 
for the $CO125$ run, from $z=7.2$ to $z=0.0$. Only a sub sample
of redshifts is shown for clarity.}
\label{fig:evolution1}
\end{figure}

\subsection{Detected shocks and Maps.}
\label{subsec:machs}

Shocks fill the simulated volume in a very complex way
(e.g. Miniati et al.2001, Ryu et al.2003).
In Fig.\ref{fig:mappe} we show a 125 kpc cut of a cubic region of 
side $80Mpc$ from the {\it AD125} run at $z=0$ with post-processing
re-ionization,
reporting gas temperature and detected shocks with Mach numbers reported 
in color code.

We find that $\approx 10-20$ per cent of the cells in the simulated volume
host shocks at present epoch, with
the percentage of shocked cells increasing in denser environments.
Filamentary and sheet--like shocks pattern are usually
hosted in low density regions and at the interface of 
filaments, following the shape of the cosmic web.  
This kind of shocks follows the first infall of baryonic matter
onto accreting structures, and generates an abrupt increase in 
temperature due to the jump from a re-ionization dominated 
temperature to the gravitationally dominated one. These shocks
are commonly defined as {\it "external shocks"} (Miniati et al.2001),
and they are the strongest within the simulation, having $M>>10$.
Shocks surrounding galaxy clusters 
form spherically shaped boundaries 
at a typical distance of about $2 R_{vir}$
from the cluster center, while shocks moving inwards 
the virializing region are found more
irregular and weak, with $M < 3$. 
This kind of shock
waves is commonly defined as {\it "internal shocks"} (Miniati et al.2001).
Slightly stronger shocks (i.e. $M \sim 3$) inside $R_{vir}$ 
are episodically found in our simulations
in case of merger events.
In this case the violent relaxation due to the fluctuation of the 
gravitational potential may cause
infall of the pre--shocked gas onto the shock discontinuity 
increasing the Mach number 
(Springel \& Farrar 2007), while other kind of strong shocks 
are reverse shocks propagating trough 
the innermost regions of accreting and cold sub clumps, 
which keep themselves at the pre--shock virial temperature for several 
Gyrs during their orbiting around the center
of the main cluster (Tormen, Moscardini \& Yoshida 2004). 

An issue which is poorly addressed in the literature is the
distribution function of shocks with their Mach number. 
Fig.\ref{fig:histo} shows the Mach number distribution 
of the shocks detected in our total $145Mpc$ cubic volume at present
cosmic epoch. 
This figure also shows the effect of using
cell-to-cell velocity jumps to reconstruct the Mach number of shocks
with the VJ scheme. 
The number of stronger shocks, that are well reconstructed within 3--4 cells, 
increases with $n=1$ and this produces a flattening of the differential
distribution of shocks inside the simulated volume.
As shown in Section~6.1, using a larger number of cells to 
reconstruct the Mach number does not improve the characterization of
shocks, yet the risk becomes to mix different shock patterns and
sub-structures in the simulations.

The overall differential
distribution of shocks with their Mach number in the cosmic volume 
is very steep, with $\alpha \sim -1.6$ 
(with $M dN/dM \propto M^{\alpha}$), and the bulk of the detected shocks
at any Mach number is found in the low density regions, which
fill the majority of the volume in the simulations.
The Mach number distribution of detected shocks
becomes increasingly steeper moving towards dense environments: 
$\alpha \approx - 3$ to $-4$ is found in 
clusters and their outskirts.

The evolution with time of the differential Mach number distribution 
is given in Fig.\ref{fig:evolution1}, for
the case of a 80 Mpc cubic region extracted from the $CO125$ simulation, 
which was designed to have a suitable time--sampling 
in the analysis of outputs.
We find that before the epoch of re-ionization, $z>6$, 
roughly 30\% of the simulated volume is shocked.
Then as soon as re-ionization plays a role, the temperature of
the simulated volume increases and the Mach number
distribution of shocks at redshift $z \sim 3-6$ 
undergoes a dramatic change becoming very steep and dominated by
weak shocks.  
With decreasing redshifts, temperature in low density regions
gradually decreases and the Mach number distribution
becomes gradually flatter with the
fraction of shocked cells reaching $\sim 15$ per cent
at present epoch. 

\begin{figure*}
\includegraphics[width=0.63\textwidth]{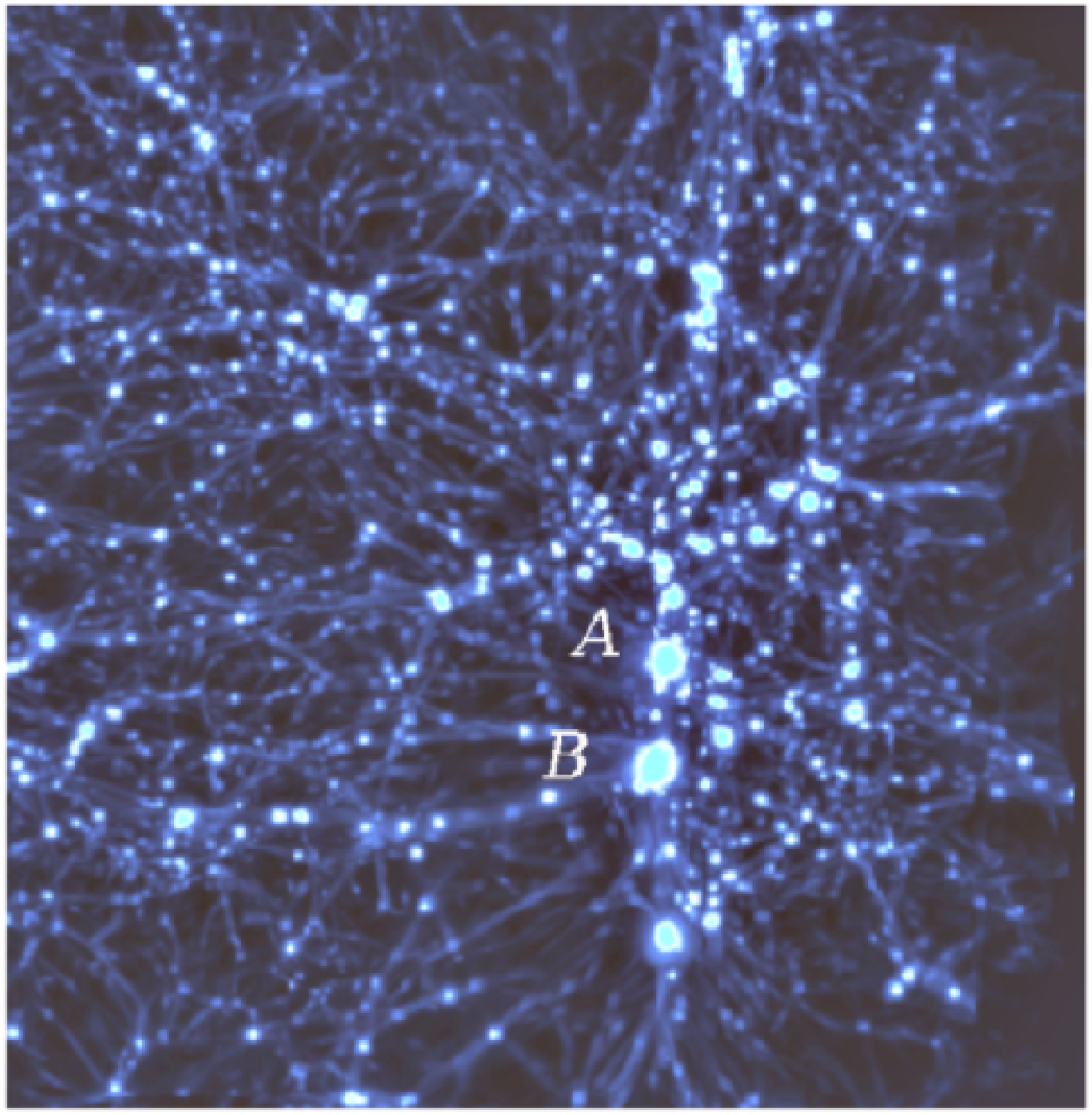}
\includegraphics[width=0.36\textwidth]{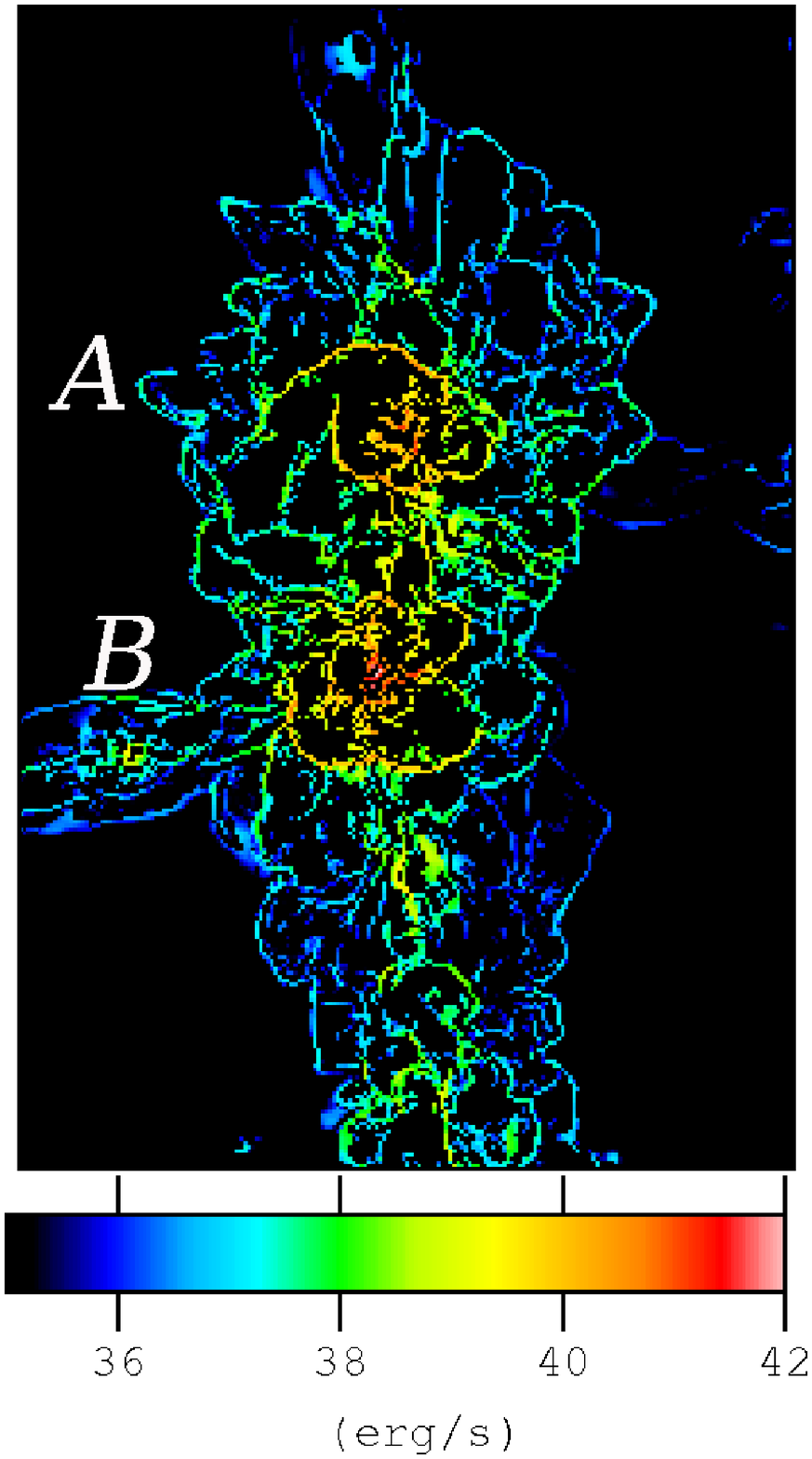}
\caption{{\it Left:}3--D rendering of baryon density for a cubic region
of side $80Mpc$, for the {\it CO125} re-simulation at $z=0$. Color coding
goes from {\it dark blue} ($\rho \sim 10^{-31} gr/cm^{3}$) to {\it pale blue}
 ($\rho > 10^{-29} gr/cm^{3}$). {\it Right:} thermalized energy flux through
shocks, for a slice of depth $125kpc$ and centered to encompass the two massive
merging clusters shown in the left panel ({\it A} and {\it B}).}.
\label{fig:flux0}
\end{figure*}

\begin{figure*}
\includegraphics[width=0.63\textwidth]{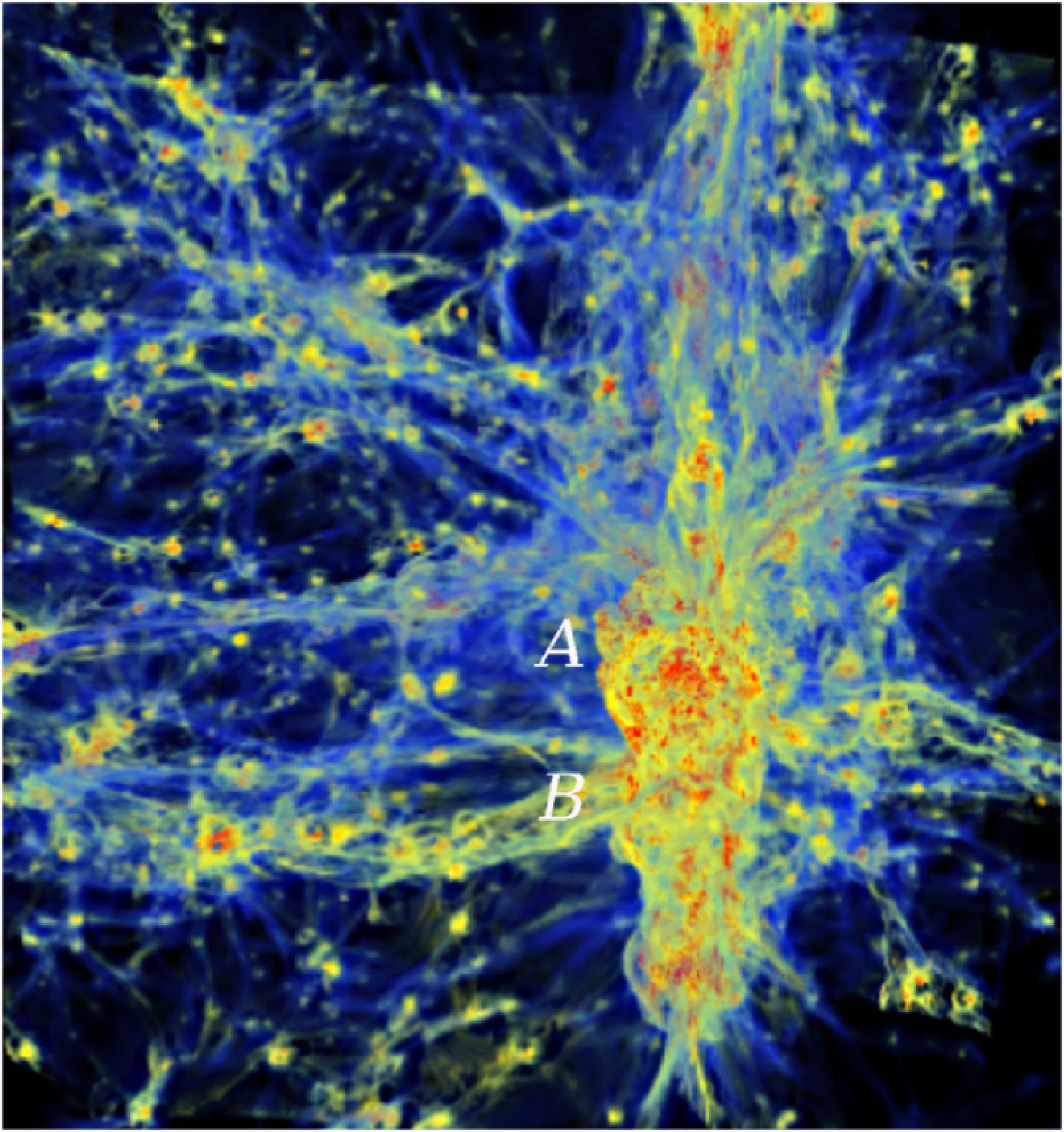}
\includegraphics[width=0.36\textwidth]{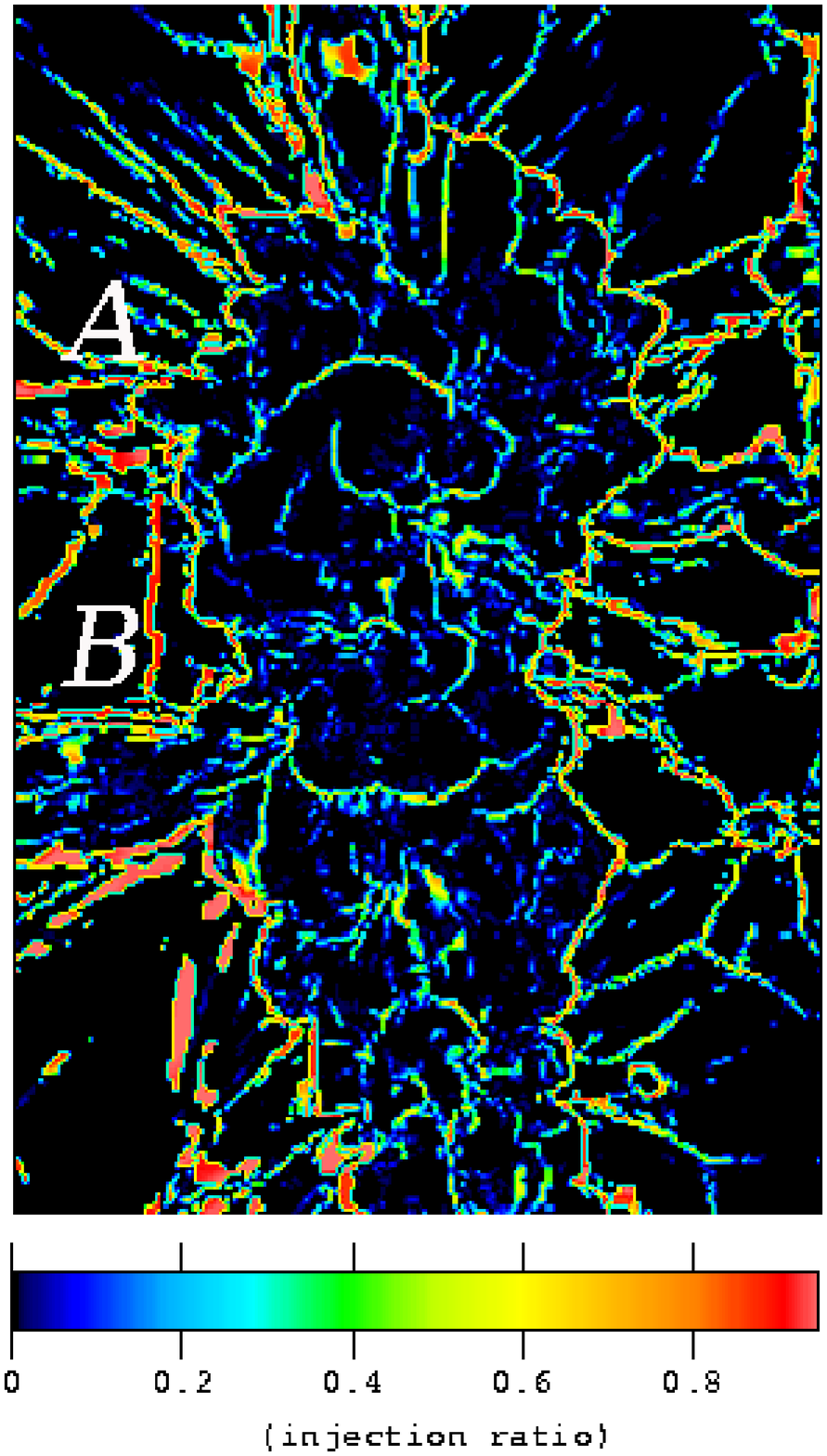}
\caption{{\it Left:}3--D rendering of the dissipated energy flux for the
same region as in Fig.\ref{fig:flux0}. Color coding goes from {\it blue} 
($f_{th}\sim 10^{33} erg/s$) to  {\it yellow} ($f_{th}\sim 10^{38} erg/s$) to
 {\it red} ($f_{th}> 10^{41} erg/s$). {\it Right:} energy ratio between 
injected CR energy flux and thermal energy flux in shock waves, for the same 
slice of right panel of Fig.\ref{fig:flux0}.}
\label{fig:flux1}
\end{figure*}

\begin{figure}
\includegraphics[width=0.48\textwidth]{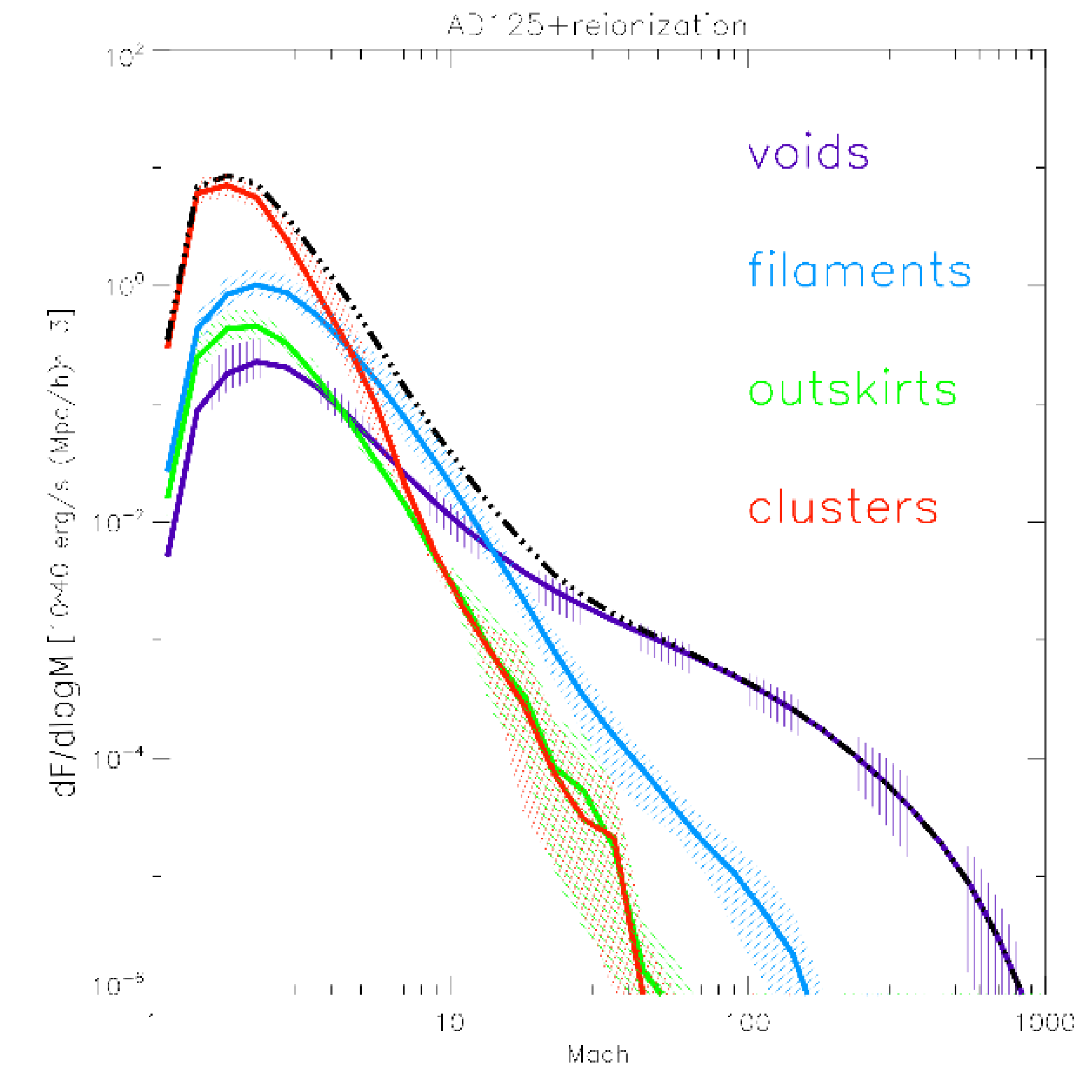}
\caption{Distribution of the thermalized energy flux 
at different over-densities, 
for the whole {\it AD125} and normalized to a comoving volume 
of $(1Mpc/h)^{3}$. The shadowed regions show the cosmic variance within
our 6 simulations, the dot-dashed line shows the global average within
the sample.}
\label{fig:histo_flux}
\end{figure}

\subsection{Energy Flux and thermalised energy}
\label{subsec:thermal}

The energy flux converted into thermal energy of the gas at a shock 
is given by the Rankine--Hugoniot jumps conditions, which
relate the flux of the kinetic energy crossing 
the shock, $E_{kin}$, and the resulting
thermal flux in the post--shock region, $f_{th}$. 
This relation can be expressed by means 
of a simple $\delta(M)$ parameter (e.g. Ryu et al.2003):

\begin{equation}
\delta(M)= f_{th}/f_{\phi}=
v_{2} \left[
{{E_{th,2}}\over{E_{kin,1}}} - \left({{\rho_{2}}\over{\rho_{1}}} 
\right)^{\Gamma} \right]
\label{eq:eq_flux}
\end{equation}

where $E_{th,1}$ and $E_{th,2}$ are the thermal energies
in the pre-- and post--shock regions, $E_{kin,1}$ is the
kinetic energy of the shock, 
and $\Gamma$ is the adiabatic exponent ($\Gamma =5/3$).
It is useful to express $\delta(M)$ by means of the Mach number 
(e.g. Kang et al.2007):

\begin{equation}
\delta(M) =  \frac{2}{\Gamma(\Gamma-1)M^{2}R}
\left[ \frac{2\Gamma M^{2}-\Gamma +1}{\Gamma +1} -R^{\Gamma} \right]          
\label{eq:eq_delta}
\end{equation}

where $R$ is the density compression factor:

\begin{equation}
R = {{\rho_{2}}\over{\rho_{1}}} = {{\Gamma +1 }\over{ \Gamma -1 + 2/M^{2}}}
\end{equation}

We notice that 
Eq.\ref{eq:eq_flux} strictly holds only 
in case of a negligible CR energy density, 
otherwise the feedback of these CR 
on the shock itself is expected to severely decrease the
efficiency of thermalisation of the kinetic energy flux 
(see next Section).

Fig.\ref{fig:flux0} (Right panel)
shows a 2--dimensional cut, with depth=125 kpc, of 
the measured thermal energy flux in shocked
cells, at the present epoch
and for a region centered around two massive 
($M\sim 4\cdot 10^{14}M_{\odot}$ and 
$M \sim 10^{15}M_{\odot}$) galaxy clusters. These clusters
belong to a large scale filament (see Left panel of Fig.\ref{fig:flux0}),
for which we provide also a 3--dimensional rendering of the 
thermal energy flux through shocks (Left panel
of Fig.\ref{fig:flux1})\footnote {We generated 3--dimensional 
distribution of data by means of the visualization tool VISIVO 
(Comparato et al.2007, http://visivo.cineca.it)}. 

The differential distribution of the
thermalised flux as a function of the Mach number of the 
shocks is reported in Fig.\ref{fig:histo_flux}. 
This shows the differential distribution calculated 
for the total volume
at the present epoch and 
normalized to the comoving volume of $(1Mpc/h)^{3}$.
Solid lines give average values,
while dashed shadows give the variance spanned by the six 
different $80Mpc$ cubic sub samples of the AD125 simulation. 
The variance in the distribution 
is fairly small, $\sim 30$ per cent 
at the peak, although this increases for the stronger shocks that are 
rare. 

\noindent
We find that the 
total processed thermal energy across
cosmological shocks in our simulations is 
$f_{th} \approx 4 \cdot 10^{47} ergs/s$ at the present epoch.
This is of the same order of magnitude of the value of the
total processed thermal energy found by 
Ryu et al.(2003) and by Pfrommer et al.(2006),
for the same $\approx$145 Mpc cubic volume.
However, as discussed in Sect.\ref{subsec:gc} the deficit in
massive halos in our clusters sample 
may induce a slightly smaller level of thermalised energy flux in the
volume.

\noindent 
For Mach numbers $< 20$ (which provide about the 99 percent of
the total thermal flux in the simulated volume) 
the distribution in Fig.\ref{fig:histo_flux} has 
$\alpha_{th}\approx -2.7$ (with $f_{th}(M)M \propto M^{\alpha_{th}}$), 
and is steeper than that in Ryu et al.(2003), $\alpha_{th}\approx -2$,
while is consistent with that in Pfrommer et al. (2006), $\alpha_{th}\approx
-2.5$.

\noindent
We find that $\approx 70$ per cent of the total thermal 
energy flux dissipated at shocks comes
from the virial region of galaxy clusters (because of their large
matter density) and that the bulk of the thermalisation happens
at shocks with $M \approx 2$ (Fig.\ref{fig:histo_flux}). 
These relatively weak shocks are also responsible for the bulk
of the thermalisation in lower density environments, 
although stronger shocks may provide a sizable 
contribution in these regions.

\begin{figure}
\includegraphics[width=0.48\textwidth]{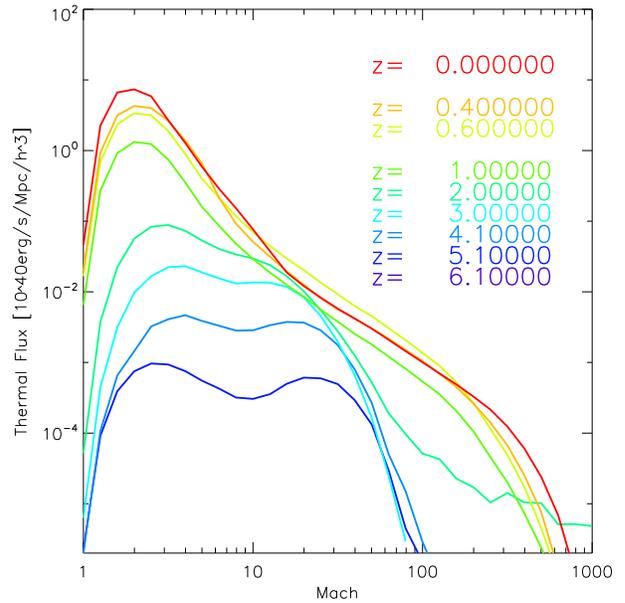}
\caption{Time evolution of the distribution of thermal energy flux
at shocks for the same volume as in Fig.\ref{fig:evolution1}, 
from $z=6.1$ to $z=0.0$. Only a sub sample
of redshifts is shown for clarity.}
\label{fig:flux_evolution}
\end{figure}

The time evolution of the distribution of the thermal energy
dissipated at shocks as a function of the shock-Mach number
is an important issue.
A relevant example is reported in 
Fig.\ref{fig:flux_evolution} that is obtained for the same 
volume of Fig.~\ref{fig:evolution1} ($CO125$).
The evolution of the 
overall distribution follows a similar behavior with cosmic
time found for the number distribution of shocks, 
with strong shocks becoming
more frequent at evolved times when the energy density of the
background becomes lower (see also Pfrommer et al. 2006).
The integrated (over cosmic time) thermal energy dissipated at shocks in our
$(145 {\rm Mpc})^{3}$ volume is  
$E_{TH} \approx 2 \cdot 10^{64} ergs$, which is consistent  with 
the values reported in Pfrommer et al. (2006) and Ryu et al.(2003), 
also by taking into account the deficit in the halos mass 
in our simulations (Sect.\ref{subsec:gc}).

\subsection{Acceleration of Cosmic Rays}
\label{subsec:CR}

The injection and acceleration of Cosmic Rays at shocks is a complex 
process.
It is customary to describe the acceleration according to the
diffusive shock acceleration (DSA) theory (e.g. Drury \& Voelk 1981; 
Blandford \& Ostriker 1978). This theory applies when particles 
can be described by a diffusion--convection equation across the shock.
There is some general agreement on the fact that 
strong shocks may channel a substantial fraction
of their energy flux into the acceleration of CR which in turn
should back react modifying the structure of shocks themselves.
Recent advances rely on the theory of non linear
shock acceleration, which describes the acceleration
of CR in shocks whose structure is modified by the back--reaction
of CR energy (e.g., Ellison, Baring, Jones 1995;
Malkov 1997; Kang, Jones \& Gieseler 2002; Blasi 2002, 2004a; 
Kang \& Jones 2005; Amato \& Blasi 2006).
The most relevant uncertainty in the description of the particle
acceleration at these shocks is the injection model, i.e. the probability that
supra-thermal particles at a given velocity can leak upstream across the
sub shock and get injected in the CR population. 
This is because even a small variation of the injection momentum, 
$p_{inj}$, 
of supra-thermal particles produces a large difference in the estimate of
the injection efficiency at shocks (e.g. Blasi 2004b).
An other major hidden ingredient is the amplification of the magnetic
field (perpendicular component) downstream
due to CR driven instabilities and adiabatic compression, as this magnetic
field self--regulates the diffusion process of CR and supra--thermal
particles (i.e. the Larmor radius) and considerably affects the value
of $p_{inj}$.

\noindent
An additional difficulty which comes out 
is that a post--processing approach, as that followed in our paper, 
does not allow us to account for the dynamical contribution of 
CR accelerated at cosmological shocks\footnote{Attempts to model this
dynamical contribution in cosmological simulations have been recently
developed (Pfrommer et al. 2006)}. 

With all these caveats in mind, we follow the approach adopted
by Ryu et al.(2003) in which the thermalisation is calculated by means
of the standard 
Eqs.\ref{eq:eq_flux}--\ref{eq:eq_delta} and the CR 
acceleration at shocks is calculated
by making use of numerical results of non linear shock acceleration
which adopt a numerical description of the {\it thermal leakage} 
to model the injection of particles in the population of CR upstream
(Kang \& Jones 2002, KJ02). These numerical results provide an
estimate of the ratio between the energy flux trough a shock and
the energy flux which is channeled into CR acceleration
at the shock by means of a simple parameter,
$\eta (M)=f_{CR}/f_{\phi}$, which depends
on the Mach number of that shock. 

Fig.\ref{fig:flux1} (Right panel) maps 
the ratio between CR and thermal energy flux 
for the same region reported in Right panel of Fig.\ref{fig:flux0}. 
This clearly shows the
role played by the Mach number in setting the level
of the injection of CR in the various environments.
Since the ratio $\eta(M)/\delta(M)$ $(=f_{CR}/f_{th})$
increases with the Mach number of the shocks, the
highest values of $f_{CR}/f_{th}$
are found in low density regions, 
at the interface layers of filaments or in
the outermost regions of galaxy clusters, where a substantial
population of relatively strong shocks is present. 
On the other hand the lower values
are typically found in galaxy clusters, where the Mach number
distribution is steep and strong shocks are rare.

\begin{figure*}
\includegraphics[width=0.46\textwidth]{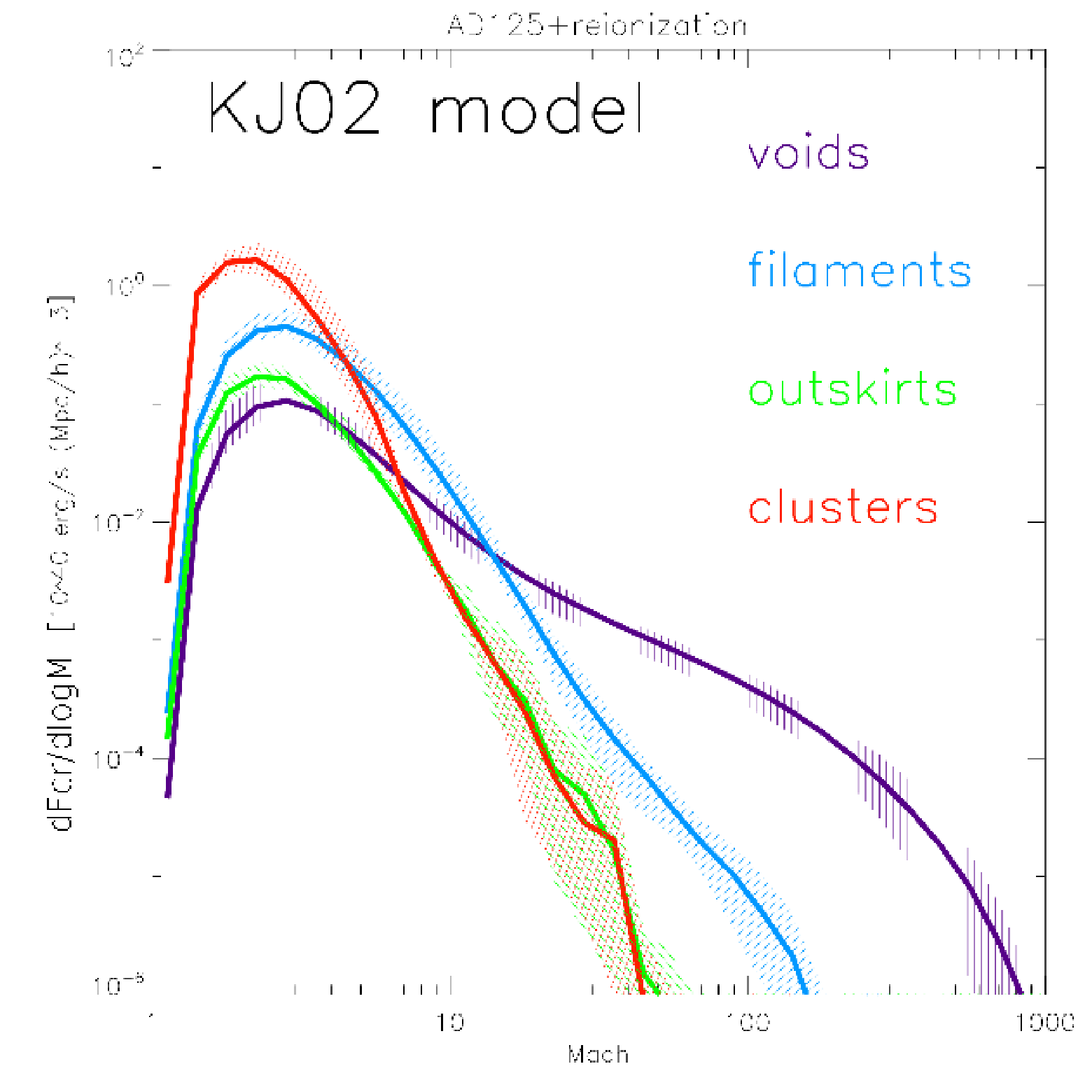}
\includegraphics[width=0.46\textwidth]{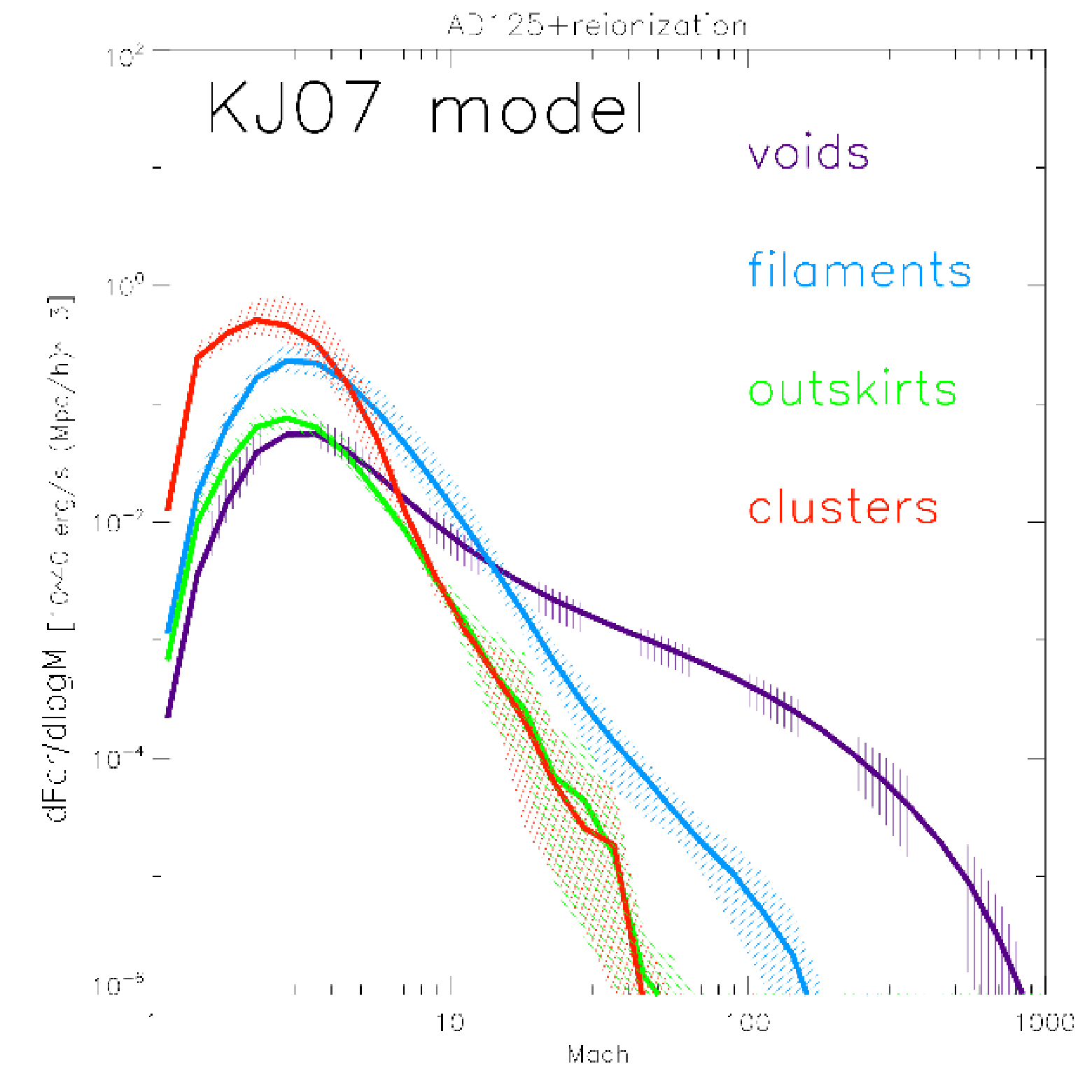}
\caption{Distribution of the injected CR flux 
at different over-densities, 
for the whole $AD125$ run with post-processing re-ionization. The 
shadowed regions show the cosmic variance within
our 6 simulations.
{\it Left} panel shows
the measured distribution according to the KJ02 recipe for the CR injection,
while {\it right} panel is for the KJ07 recipe.}
\label{fig:histo_cr}
\end{figure*}

The distribution of the energy flux injected in CR as a function of
Mach number is reported in Fig.\ref{fig:histo_cr} ({\it left} panel);
this refers to the
total simulated volume at the present epoch.
We find that the bulk of the CR acceleration at present
epoch takes place in galaxy clusters, however also filaments are expected
to contribute significantly
to the overall acceleration process.
The overall distribution has a well defined peak
which is anchored at $M \approx 2$ and has a slope 
(between $M \sim 2 - 20$)  $\alpha_{CR} \approx -2$
(with $f_{CR}(M)M \propto M^{\alpha_{CR}}$).

The value of the Mach number at the peak 
is close to (even if slightly smaller than) that 
found by Ryu et al.(2003) ($M \sim 3$), while the 
distribution is steeper than that reported in
Ryu et al.(2003) (where $\alpha_{CR} \approx -1.5$).
Since we use an approach equivalent to that
in Ryu et al.(2003) to evaluate the CR acceleration,
this difference is likely related to
our different shock detecting scheme, to our different modeling 
of the re-ionization process and also to the slightly different value of the
$\sigma_{8}$ parameter ($\sigma_{8}=0.8$ in Ryu et al.2003;
we discuss in the Appendix the influence of $\sigma_{8}$ in the 
shock statistics).
A comparison with the results in Pfrommer et al.(2006) is more
difficult since these authors use a Lagrangian Smoothed Particles Hydrodynamics
code which also include CR dynamics and a completely different approach
in the calculation of the CR injection at shocks.
The overall distribution of the energy flux injected in CR reported
in Pfrommer et al.(2006) has a slope $\alpha_{CR} \approx -1.8$ and 
is actually in between our results and those of Ryu et al. (2003).

For seek of completeness, in Fig.\ref{fig:histo_cr} (Right panel)
we also report the overall distribution of the energy flux injected in CR
by adopting the injection efficiency of CR
at modified shocks by Kang \& Jones 2007 (KJ07).
These recent calculations account for
the Alfv\'en wave drift and dissipation in the shock precursor 
and this yields a value of $\eta(M)$ which is smaller than that
adopted by Ryu et al.(2003) (at least for $M < 20$). 
As a consequence the resulting distribution of
the energy flux dissipated at shocks into CR acceleration as a function
of the shock--Mach number 
is flatter than that obtained by adopting KJ02 and $\approx$ 50 per cent less energy is expected
to be channeled into CR acceleration.

\begin{figure}
\includegraphics[width=0.48\textwidth]{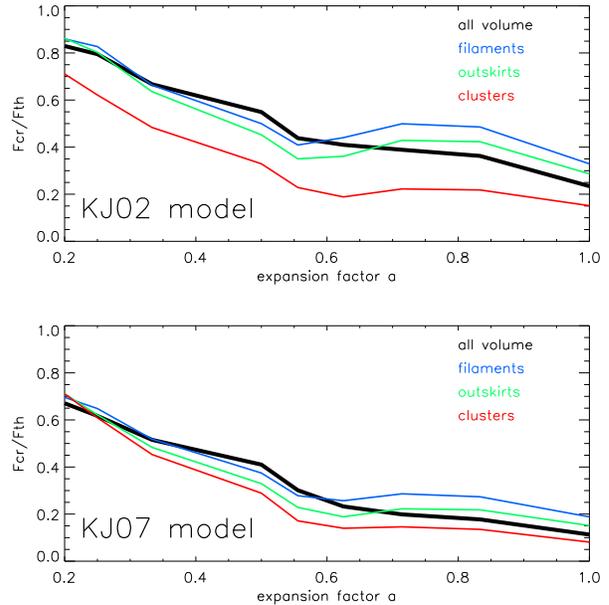}
\caption{Evolution with time of the total injection ratio $f_{CR}/f_{TH}$
for the same sub sample as in Fig.15 and \ref{fig:flux_evolution}, shown
for different environments. 
The upper panel is for the acceleration model outlined 
in Kang \& Jones (2002), while the lower panel is for the acceleration 
model of Kang \& Jones (2007).}
\label{fig:cr_int}
\end{figure}

Fig.\ref{fig:cr_int} shows the evolution 
with time of the ratio $f_{CR}/f_{th}$
for the same volume (CO125) considered in Figs.~\ref{fig:evolution1}
and \ref{fig:flux_evolution}. 
The value of $f_{CR}/f_{th}$ as measured at the
present epoch, $f_{CR}/f_{th} \sim 0.2$, is a factor $\approx$2 
smaller than that found in Ryu et al.(2003). 
By adopting the injection efficiency of CR of KJ07
the ratio $f_{CR}/f_{th}$ is even smaller, about 0.1, as the 
injection efficiency of CR in KJ07 is smaller than
in KJ02 (at least for shocks with $M < 20$).
Although Fig.~\ref{fig:cr_int} shows that CR in dense regions do
not provide a relevant back-reaction on the thermal pool during their 
acceleration (this justifies the use of Eqs.~10--12 in these 
environments), the larger values of $f_{CR}/f_{th}$ in low density 
regions and at early cosmic times suggest that following 
(run-time) the dynamics of CR and the (self consistent) non--linear
shock thermalisation and CR acceleration is mandatory in future studies 
with Eulerian--cosmological simulations.

\subsection{Shocks in Galaxy Clusters.}
\label{subsec:clusters}

In this Section we focus on the shock statistics and CR injection 
in galaxy clusters and briefly discuss their dependence on the 
cluster dynamics.
We study shocks in four representative massive galaxy
clusters extracted from the {\it AD125} simulation, at $z=0$:

\begin{itemize}

\item  {\bf C1}: a $M_{tot}\sim 7\cdot 10^{14} M_{\odot}$ cluster 
in a relaxed state;

\item  {\bf C2}: a $M_{tot}\sim 7\cdot 10^{14} M_{\odot}$ cluster subject
to an ongoing minor merger with a sub clump with mass 
$M_{tot}\sim 2\cdot 10^{13} M_{\odot}$;  

\item {\bf C3}: a $M_{tot}\sim 1\cdot 10^{15} M_{\odot}$ cluster approaching
a major merger with zero impact parameter, with a companion cluster
(with $M_{tot}\sim 4\cdot 10^{14} M_{\odot}$ ) that is at
a distance of $\sim 1.3 R_{vir}$ from the main cluster center;

\item  {\bf C4}: 
a $M_{tot}\sim 7.5\cdot 10^{14}  M_{\odot}$ cluster in a post--merging
phase (the merger occurs in the simulation $\sim 2$ Gyr in look back
time).

\end{itemize}

\begin{figure*}
\includegraphics[width=0.48\textwidth]{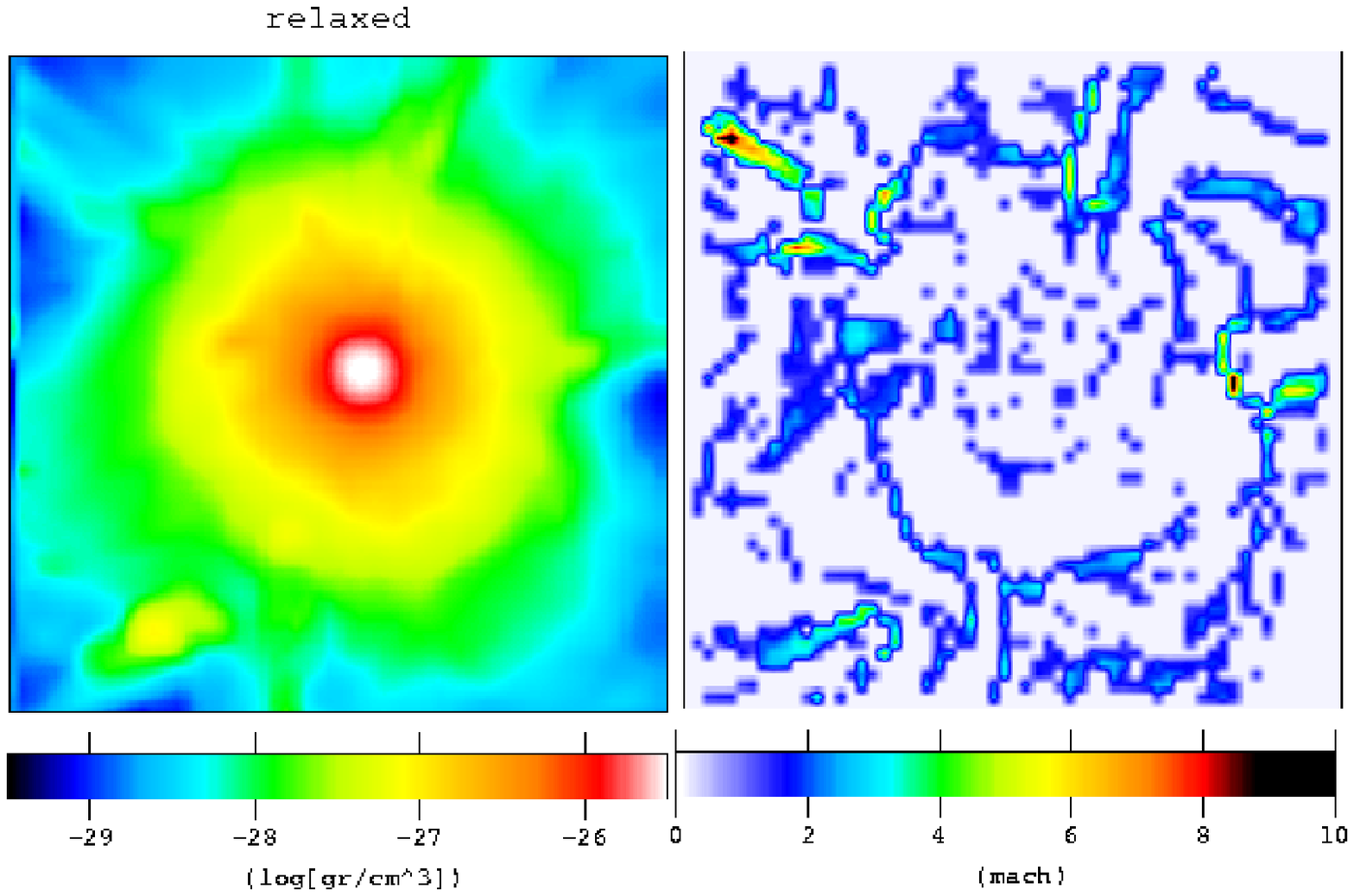}
\includegraphics[width=0.48\textwidth]{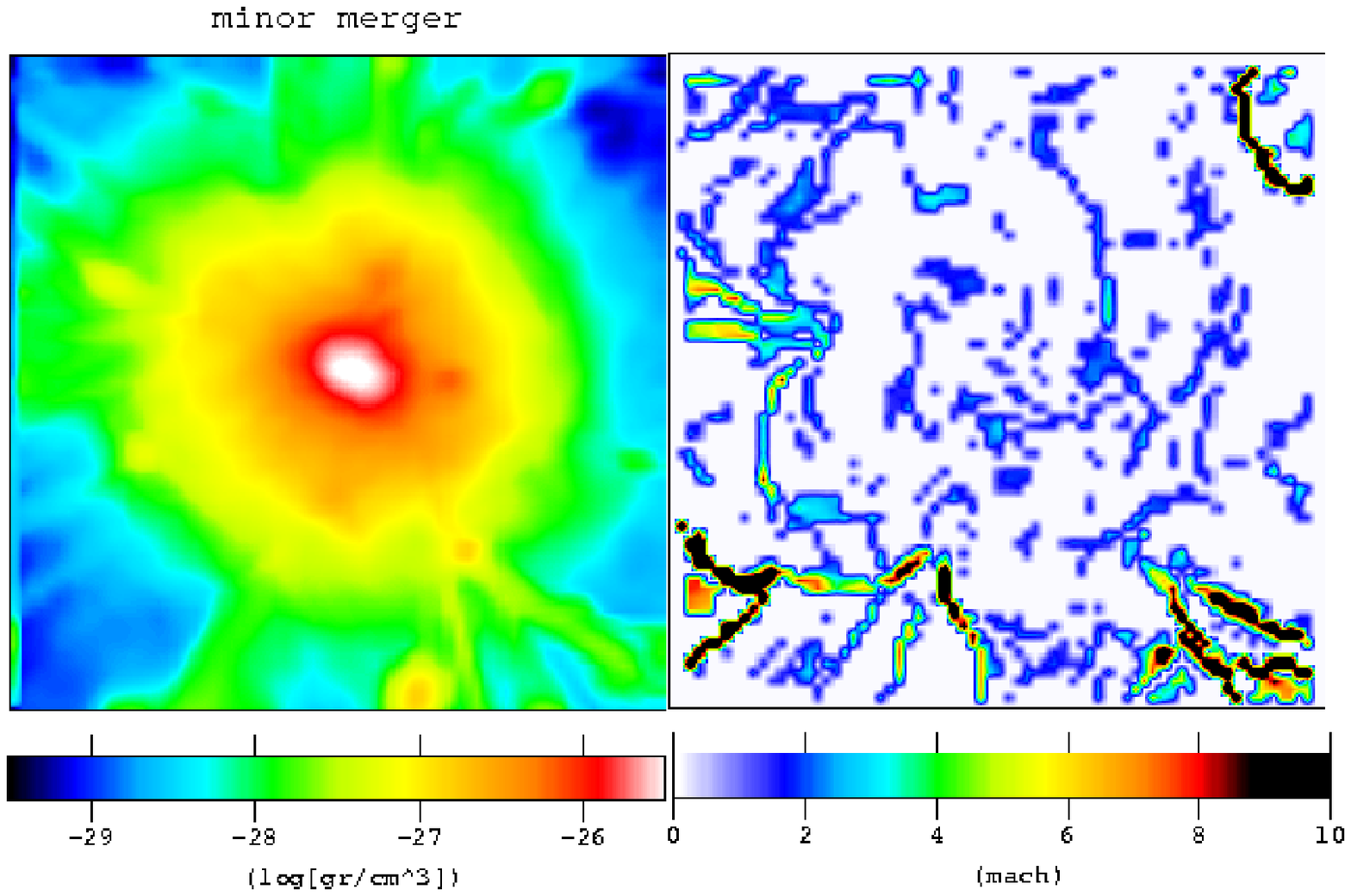}
\includegraphics[width=0.48\textwidth]{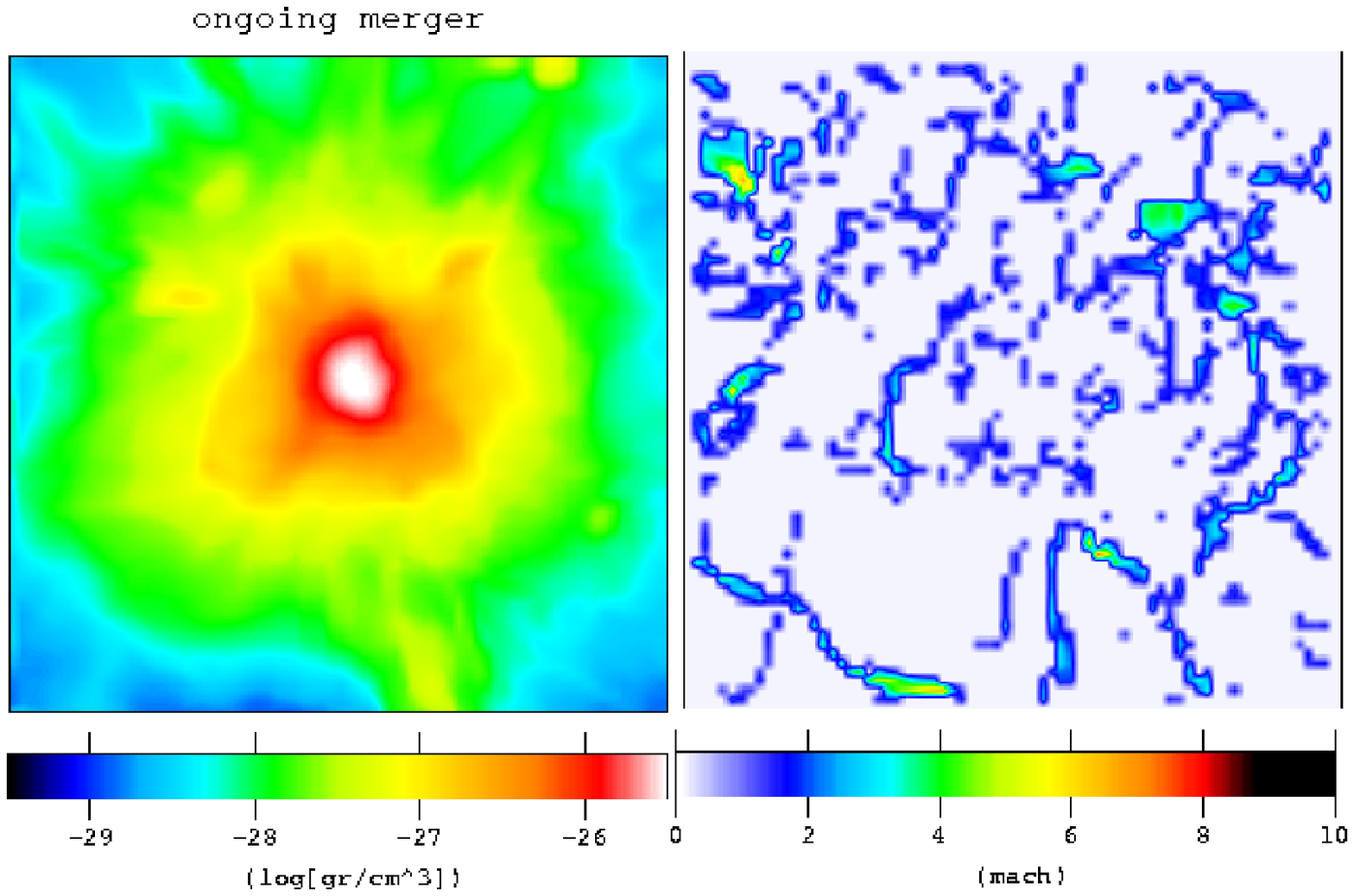}
\includegraphics[width=0.48\textwidth]{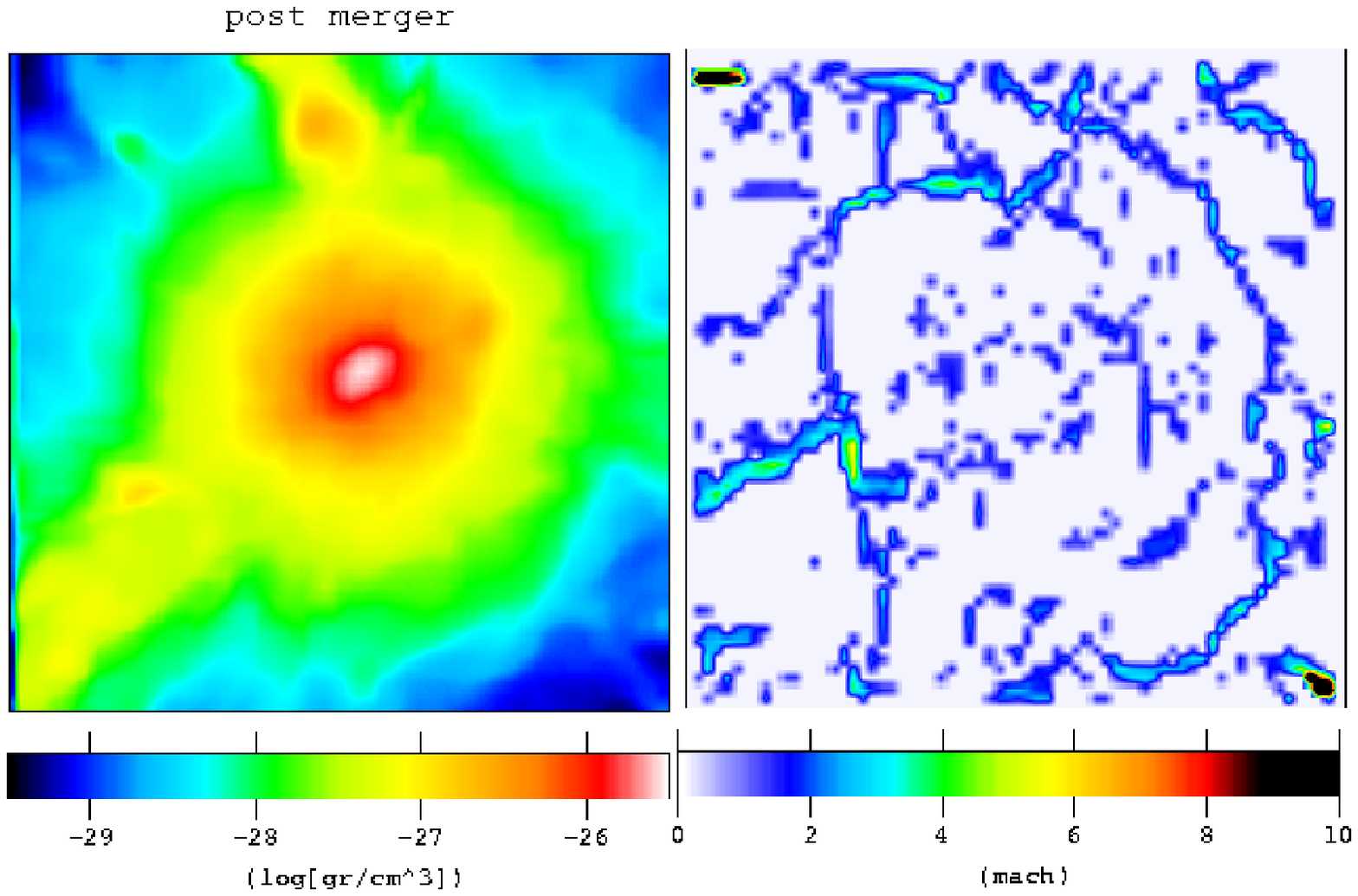}
\caption{{\it Left panels:} maps of projected baryon density
for the 4 galaxy clusters of Sect.\ref{subsec:clusters}. Every
map has a depth along the line of sight of twice the virial radius of the
correspondent cluster.{\it Right panels:} slabs
of $125$ kpc along the line of sight, showing the maps of Mach number 
for the same objects as in left panels. For displaying purposes, pixels in the images have been oversampled and convolved with a Gaussian kernel with a FWHM=2 cells.}
\label{fig:map_clusters}
\end{figure*}

\begin{figure}
\includegraphics[width=0.46\textwidth,height=0.45\textwidth]{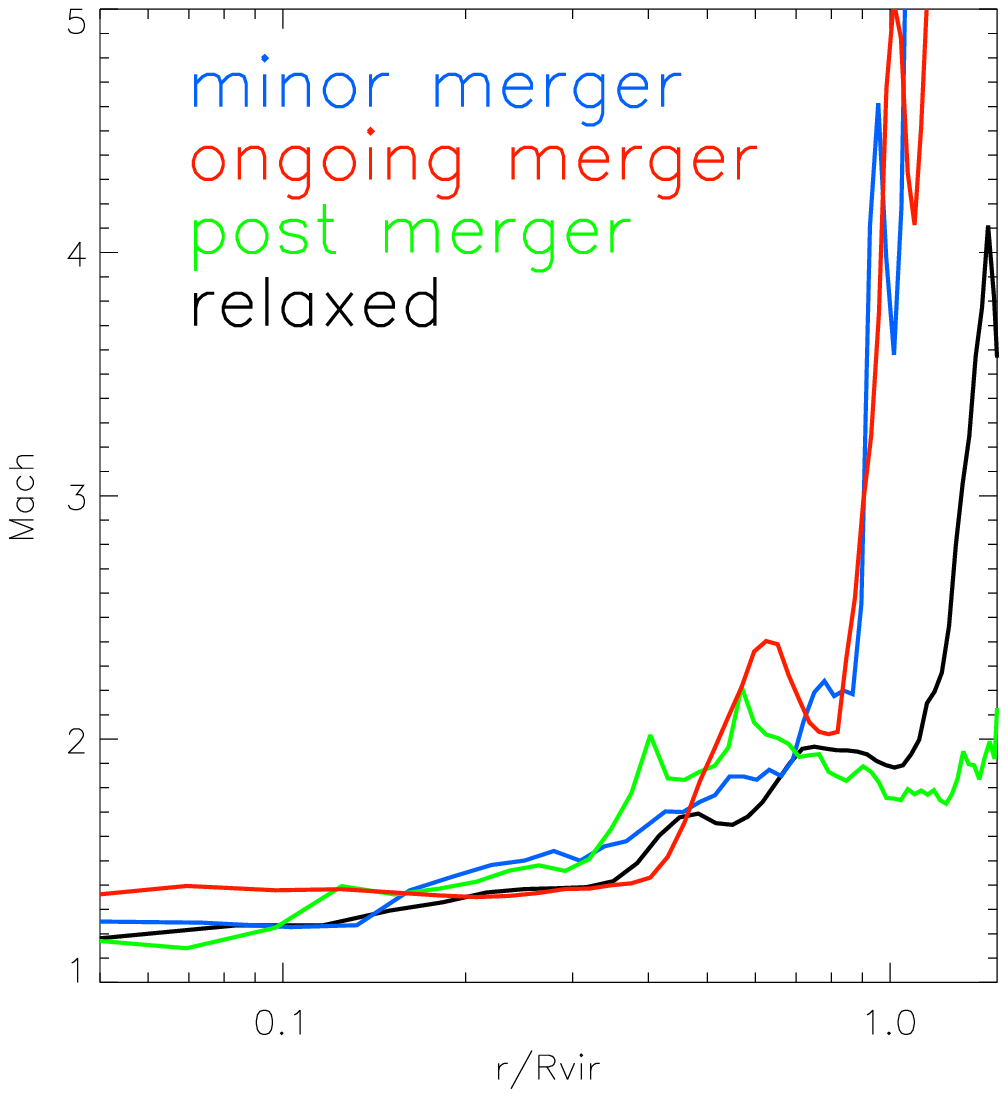}
\caption{Volume averaged profile of the mean Mach 
number of shocks for the galaxy clusters in Fig.\ref{fig:map_clusters}}
\label{fig:prof_clusters}
\end{figure}

These clusters are reported in Fig.\ref{fig:map_clusters}, that 
shows maps of  projected baryon density in a  
$(4R_{vir})^{2}$ regions centered on clusters, and maps
of the Mach number
measured in slices crossing the same regions.

\noindent
In the case of {\it C2} (minor merger) and {\it C3} (major merger) 
relatively weak, $M\approx 2-2.5$, shocks are found 
inside $R_{vir}$, while in the case of {\it C4} (post-merger) 
merger shocks have already moved outside
the internal region of the cluster, and their strength is increased as 
the ambient temperature in cluster outskirts decreases.

\begin{figure*}
\includegraphics[width=0.95\textwidth,height=0.4\textwidth]{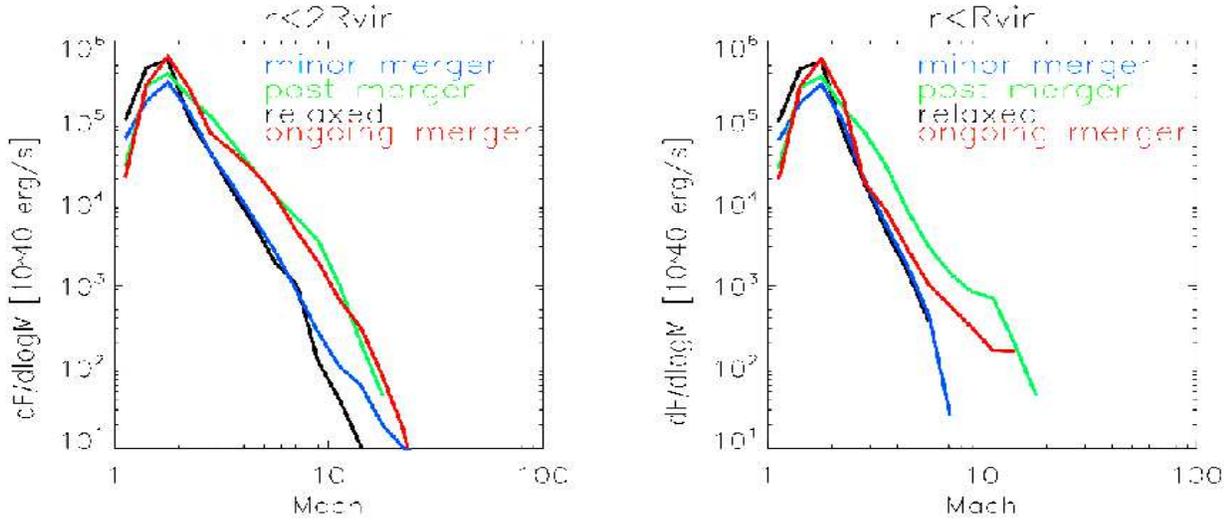}
\caption{Distribution of thermalised fluxes for the 
galaxy clusters presented in the text. 
Distribution are normalized for the volume
of the most massive one, and are taken from spheres of radius $=2R_{vir}$ ({\it left})and 
$R_{vir}$ ({\it right}) around each cluster.} 
\label{fig:clust_distrib}
\end{figure*}

The volume averaged Mach number of shocks in the four galaxy clusters 
is reported as a function of distance from cluster centers in 
Fig.\ref{fig:prof_clusters}.
Within the virial radius shocks are weak in line with
expectations from semi--analytical models that indeed found 
shocks with $M>3$ extremely rare in galaxy clusters 
(Gabici \& Blasi 2003). This is also highlighted in 
Fig.\ref{fig:clust_distrib}, where we show
the distribution of the thermal flux dissipated at shocks with shock-Mach 
number; distributions in different clusters are reported 
normalized to the volume of the most massive cluster.
All distributions are steep, with differences from cluster to cluster due 
to the effect of the dynamical state of the clusters.
Inside $R_{vir}$ {\it C1}, {\it C2} and {\it C3} have similar distributions, 
while {\it C4} shows some excess of rare shocked cells with $M \approx 3-7$.
On the other hand, an excess of shocked cells with $M \approx 3-10$
is found in the external regions of {\it C3} and {\it C4}.

\noindent
Also in the case of clusters our 
distributions of the thermalised energy flux
are steeper than those reported in
other similar works: 
we find $\alpha_{th} \approx - 4$ to $-5$, 
while $\alpha_{th} \approx -3$ to $-4$ is obtained 
by Pfrommer et al.(2007), where the Lagrangian
SPH code Gadget--2 was employed.

\begin{figure*}
\includegraphics[width=0.95\textwidth,height=0.4\textwidth]{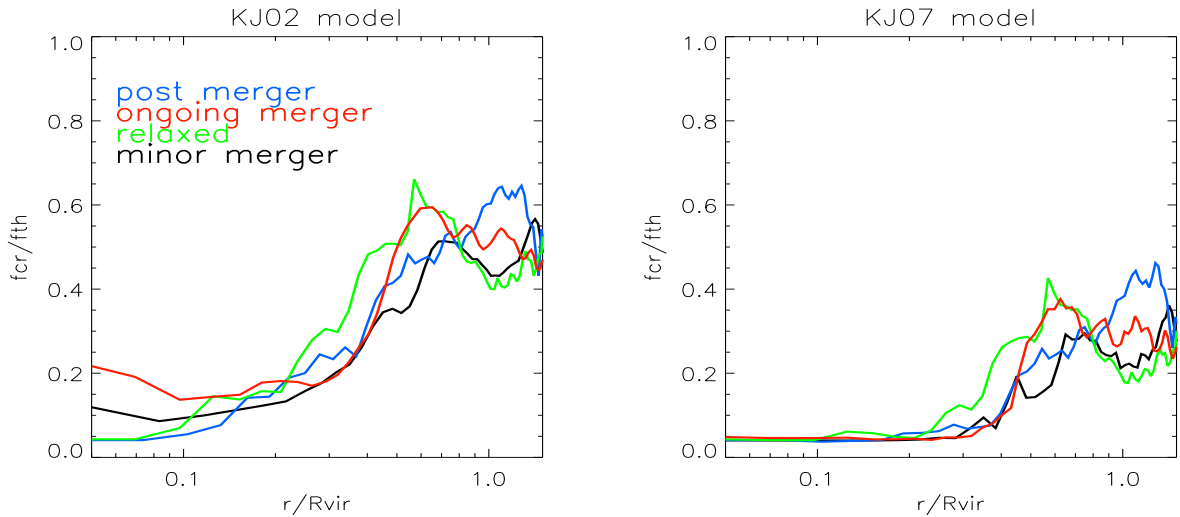}
\caption{Volume averaged profiles of $f_{CR}/f_{th}$ for the four galaxy clusters. {\it Left} panel 
is for the KJ02 model and {\it right} panel is for the KJ07 model.} 
\label{fig:prof_cr_clusters}
\end{figure*}

The radial profile of the ratio $f_{CR}/f_{th}$ in our
clusters is reported in Fig.\ref{fig:prof_cr_clusters}.
Here we show the results in the case of both the KJ02 ({\it left} panel) 
and KJ07 ({\it right} panel) models.
Inside the virial radius we do not find any relevant difference 
between our clusters. 
This is because, independently of the dynamical status of the clusters,
the bulk of the energy is dissipated in the CR acceleration and
in the thermal component at relatively weak shocks.
The maximum value of $f_{CR}/f_{th}$ is found at distance $\geq R_{vir}$
from the cluster center, $f_{CR}/f_{th} \approx 0.5$ and 0.3
using the KJ02 and KJ07 model, respectively.

\section{Discussion and Conclusion.}
\label{sec:conclusions}

In this paper we have investigated the
properties of shocks in cosmological numerical simulations.
This subject is particularly intriguing as 
shock waves propagating trough LSS are the responsible for
the heating of the ICM and may be 
important sources of CR in the Universe.
This subject has been already investigated in several 
papers under different numerical approaches 
(Miniati et al.2000; Miniati et al.2001; Keshet et al.2003;
Ryu et al.2003; Pfrommer et al.2006; Kang et al.2007; Pfrommer et al.2007{\footnote{See also Skillman et al.2008, which appeared after this paper was submitted.}}).

We study shock waves by means of a post processing
procedure. Although 
this is similar to Ryu et al.(2003) and Kang et al.(2007),
our approach differs from previous ones in several points.
First we use a different numerical codes, the public
version of ENZO (e.g. Bryan \& Norman 1997), to simulate LSS 
(Sec.\ref{sec:enzo}).
Second we adopt a more appropriate treatment of
the re-ionization in our simulations (Sec.\ref{sec:reionization})
and third we use a different approach to
catch shocks in our simulations and to measure their strength
(VJ method, Sect.\ref{subsec:vj}).

\subsection{Results}

We simulated a large cosmic volume, 
$(145 {\rm Mpc})^{3} \approx (103 {\rm Mpc/h})^{3}$, 
with a fixed grid resolution
of 125 kpc. Additional simulations were designed and used
to investigate the effect of spatial resolutions and
of the $\sigma_{8}$ parameter (see Appendix).

\noindent
In the following we summarize the main results obtained :

\begin{itemize}

\item {\it Re-ionization}: 
in Sect.\ref{subsec:reion} we have shown that 
a correct treatment of the re-ionization 
is crucial to have a viable description of shocks with a post
processing procedure.
In particular using a constant temperature
floor in the simulated data at a fixed redshift may cause a
somewhat artificial flattening (or pile up) of the 
differential distribution of the shocks 
with Mach number.
This is because the velocity of the sound is artificially boosted
up in low density and cold regions.
To overcome this point we derive formulas giving the 
typical temperature of the gas as a function of the local density 
by fitting data obtained from
simulations which include a specific modeling of the
re-ionization in run--time.
These formulas are found to be consistent with 
Katz et al.(1996) and Valageas et al.(2002) 
and can be used to model the temperature background of
adiabatic simulations in a post processing procedure.
In Section 4 (Fig.\ref{fig:madau-vs-post}) we have shown 
that our post processing procedure
is indeed consistent with the results from simulations
with run--time re-ionization.

\item {\it Methods to derive the Mach number of shocks}:
in Sect.\ref{subsec:Ryu0} and \ref{subsec:vj}  we have discussed in some 
detail two different methods to catch shocks in simulated data and 
to estimate their Mach number.
These methods are the temperature jump (TJ) and the
velocity jump (VJ) and rely with the jumps in temperature and
velocity across shocked cells, respectively.

The shock discontinuity is typically spread over a few cells
and the risk in measuring the Mach number of shocks trough cell-to-cell
velocity or temperature jumps (or jumps across a few cells) is to
underestimate the Mach number of shocks.
To study this point in the context of our numerical simulations
we perform several shock--tube tests with ENZO
and calculate maps of shocks in our simulations under different
approaches (Sect.~6.1). We conclude that shocks in our simulations
are best characterized from velocity (VJ) or temperature (TJ) jumps
measured across three cells.

Both the VJ and TJ schemes use ideal conditions across non shocked cells
(i.e. no velocity and no temperature gradients) and this may cause
major uncertainties in the characterization of the shocks
in a post processing procedure (Sect.\ref{subsec:tjun}--\ref{subsec:vjun}).
This is because the velocity field and temperature distributions in the
cosmological data sets are very complex and the passage of 
a shock establishes thermodynamical gradients which are superimposed
to already existing ones.
We discuss the strength of the uncertainties
on the value of the Mach number from the two schemes by means of  
Monte Carlo extractions of temperature and velocity variations 
across non shocked cells in our data sets, and 
show that the VJ method may be more reliable, at least in the case of
weak shocks and especially in low density environments (Figs.8--9).

\noindent
Besides these uncertainties we find that the two methods 
yield statistically similar Mach number distributions of shocked
cells in our simulations (Fig.~12) suggesting that 
the characterization of shocks in our simulations is fairly solid;
in Sect.7 we adopt here the VJ method.

\item 
{\it Morphology of LS shocks}: 
in Sect.\ref{subsec:machs} we discuss the morphology
of the shock--patterns detected in our simulated data sets.
About $15$ per cent of the cells in our simulations are found to
host shocks at the present epoch, and this fraction slightly
decreases with look back time for post--reionization epochs.
In qualitative agreement with previous studies 
(Ryu et al. 2003; Pfrommer et al. 2006) we find that these shocked cells form
spectacular and complex patterns associated with the cosmic web. 
Filamentary or sheet--like shocks are found outside the virial regions
of clusters and around filaments, while more regular spherical
structures surround galaxy clusters.

\item
{\it Number Distributions of LS shocks}: 
we study 
the number distribution of shocked cells as a function of the
Mach number of the shocks.
An important point here is that thanks to the Eulerian scheme
of the ENZO code we were able to follow the 
hydrodynamics of the LS shocks also in very low density regions, 
whose exploration is challenging for present Lagrangian schemes.

We find that 
the bulk of cosmological shocks is essentially made by weak $M \leq
2$ shocks and that the number distribution of shocks
can be grossly described by a steep power law $M dN/dM \propto M^{\alpha}$.
When considering the Mach number distribution of shocked cells
in the total simulated volume 
we find an overall steep distribution $\alpha \approx -1.6 $
which is dominated by the contribution from voids and filaments.
This distribution steepens with increasing the cosmic over-density
and becomes very steep ($\alpha \approx -3$ to $-4$) in 
the case of galaxy clusters 
demonstrating the increasing rarity of strong shocks in these denser
(and hotter) regions, where basically $M<3$.

\item {\it Energy dissipated at LS shocks} :
the energy dissipation at LS shocks is the main focus of the
previous studies on this topic (e.g., Miniati et al.2001; Keshet et al.2003; 
Ryu et al.2003; Pfrommer 
et al. 2006; Kang et al.2007; Pfrommer et al.2007).
Following Ryu et al.(2003) we calculate the energy rate dissipated
in the form of thermal energy at shocks by means of hydrodynamical
jump conditions (Eq.\ref{eq:eq_delta}).
In agreement with these previous studies we find that 
about $\sim 4 \cdot 10^{47} erg/s$ are dissipated at shocks 
in a $(103 Mpc/h)^3$
volume in the simulations at the present epoch.
The bulk of the energy in our simulations is dissipated 
in galaxy clusters which indeed
contribute to about $75$ per cent of the total energy dissipation (about
$80$ per cent if we also include the contribute from cluster outskirts),
 while filaments contribute
to a $15$ per cent of the total energy dissipation. 
We calculate the distribution of the energy flux dissipated at LS
shocks with shock-Mach number.
When considering shocked cells
in the total volume we find that the distribution
is steep, $\alpha_{th} \approx -2.7$ ($f_{th}(M)M \propto M^{\alpha_{th}}$)
and peaks at $M \approx 2$.
Although in qualitative agreement with previous studies,
our distribution is steeper than that obtained by Ryu et al.(2003) 
that also used cosmological simulations based on a Eulerian scheme.
This difference is mostly due to 
our more complex treatment of the re-ionization background and  
to the use of the VJ scheme to measure the Mach number of shocks.

Following Ryu et al.(2003)
we calculate the efficiency of the injection of CR at LS shocks
according to Kang \& Jones (2002).
We obtained Mach number distributions of the energy flux dissipated
into CR acceleration in our simulated data sets.
Our results are in line with previous findings,
although our distributions are 
steeper than those in Ryu et al. and
slightly steeper than those in Pfrommer et al.(2006 \& 2007).

In agreement with Pfrommer et al. we find that the bulk of
the energy dissipated in the form of CR at shocks 
is shared between clusters and filaments so that CR--acceleration
happens in regions broader than those where thermal energy is
dissipated at shocks and the ratio between CR energy and
thermal energy increases in lower density regions (Fig.\ref{fig:flux1}).
When considering all the shocked cells in our simulations 
we find that the ratio between the energy dissipated in the
form of CR-acceleration and of thermal energy 
at present epoch is $f_{CR}/f_{th} \approx 0.2$ and this ratio
becomes smaller in galaxy clusters.
These fractions are a factor $\approx$2 smaller than those found by Ryu et al.
and this is likely due to the steeper distributions that
we obtained in our analysis.

\item {\it Galaxy Clusters}:
in Sect.\ref{subsec:clusters} we briefly discuss the case 
of shocks propagating in galaxy
clusters. We find very steep distributions 
of the number of shocks and of the energy dissipated at shocks
as a function of the shock-Mach number.
The typical Mach number within the virial radius is $M \approx 1.5$,
in nice agreement with semi--analytical studies which are
indeed appropriate for virialised systems (Gabici \& Blasi 2003).
At larger distance from the cluster center stronger shocks
are found and their presence is strongly correlated with
the dynamical status of the clusters.
Remarkably the rarity of moderate--strong shocks in the cluster
central regions (within $\approx$ Mpc distance from cluster center)
makes the ratio $f_{CR}/f_{th}$ very small, expecially when the Kang \& Jones
(2007) model for the injection of CR at shocks is adopted (Fig.~25).
\end{itemize}

\subsection{Comment on the injection of CR}

The use of numerical simulations is mandatory 
to understand the properties of LS shocks and to investigate their
role in the injection of CR in LSS.
Although we use a different approach
with respect to other studies, 
our findings for the energy dissipated in the form of
CR at these shocks are grossly consistent with previous studies
(Ryu et al. 2003; Pfrommer et al. 2006).

However, the astrophysical problem is extremely complex and 
several {\it hidden} ingredients in the adopted procedures 
are potentially sources of large uncertainties.
As discussed in Sect.\ref{subsec:CR} the efficiency of CR acceleration 
at shocks is investigated following several approaches.
In the present paper we have used the acceleration efficiency 
resulting from numerical calculations of modified shocks
(following Ryu et al. 2003 and references therein).
On the other hand Pfrommer et al.(2006) use a linear theory 
with the efficiency modified to account for
saturation effects at large values of the Mach numbers 
(actually to limit the CR efficiency at $\approx 50$ per cent).
These two approaches are formally {\it radically} different, 
but nevertheless they 
provide an overall estimate of the CR injection efficiency 
which is not dramatically different in the case of 
typical shocks with $M \approx 2-4$.
The main {\it hidden} ingredient in 
the efficiency of CR acceleration comes from 
the commonly adopted {\it thermal leakage injection} scenario  
which essentially adopts as minimum momentum of the particles that take part
in the acceleration process, $p_{inj}$, a multiple of the
momentum of the thermal particles, $p_{inj} = x_{i} p_{th}$.
The choice of $x_{i}$ is a {\it guess}, since this depends on physical
details which are still poorly known (e.g., Blasi 2004).
In Ryu et al.(2003) (and thus in our paper) the fraction of protons
injected into the CR population at shocks is taken of the
order of $\approx 10^{-3}$ which is not far from (even if larger than)
the resulting efficiency from the 
assumption of $p_{inj} \approx 3.5 p_{th}$ adopted in 
Pfrommer et al.(2006).
Although this parameter is somewhat constrained by the theory
(e.g., Malkov 1998), it should be stressed that 
having a slightly different value of $x_{i}$ (e.g. $x_{i}$=3.8 instead of 3.5)
would have the net effect to reduce the
acceleration efficiency by nearly one order of magnitude.

\subsection{Constraints from observations}

Theoretical arguments suggest that the bulk of CR in galaxy clusters
should be in the form of supra--thermal protons (e.g. Blasi, Gabici,
Brunetti 2007 for a recent review).
Gamma ray observations of a few nearby galaxy clusters
limit the energy density
of CR protons in these clusters to $\approx$ 20 per cent 
of the thermal energy (Pfrommer \& Ensslin 2004; Reimer 2004).
Although a detailed comparison with these limits would require 
to follow the advection and accumulation process of CR in clusters
during the simulations (e.g. Miniati 2003; Pfrommer et al. 2006), 
the level of 
injection rate of CR that we found in clusters and cluster outskirts 
(Sect.\ref{subsec:clusters}) allows us to reasonably conclude 
that our results are barely consistent with these limits.

On the other hand more stringent upper limits can be obtained from
present radio observations. The bulk of galaxy clusters does not
show evidence of extended Mpc--scale synchrotron radio emission and 
this can be used to constrain the population of secondary electrons and 
thus that of the primary CR protons from which these
secondaries would be injected (Brunetti et al. 2007).
These limits are very stringent and actually represent
a challenge for simulations:
in the case that the ICM is magnetized at 
$\approx \mu$G level (consistent with Rotation Measures, e.g. Govoni 
\& Feretti 2004) the energy of CR should be at $\leq$ few percent
of the thermal energy (when the spectrum of CR is fixed at that
expected from simulations, i.e. $s\approx 2 - 2.5$, $N(p)\propto p^{-s}$ for
$M\geq 3$).

\begin{figure*}
\includegraphics[width=0.95\textwidth,height=0.8\textwidth]{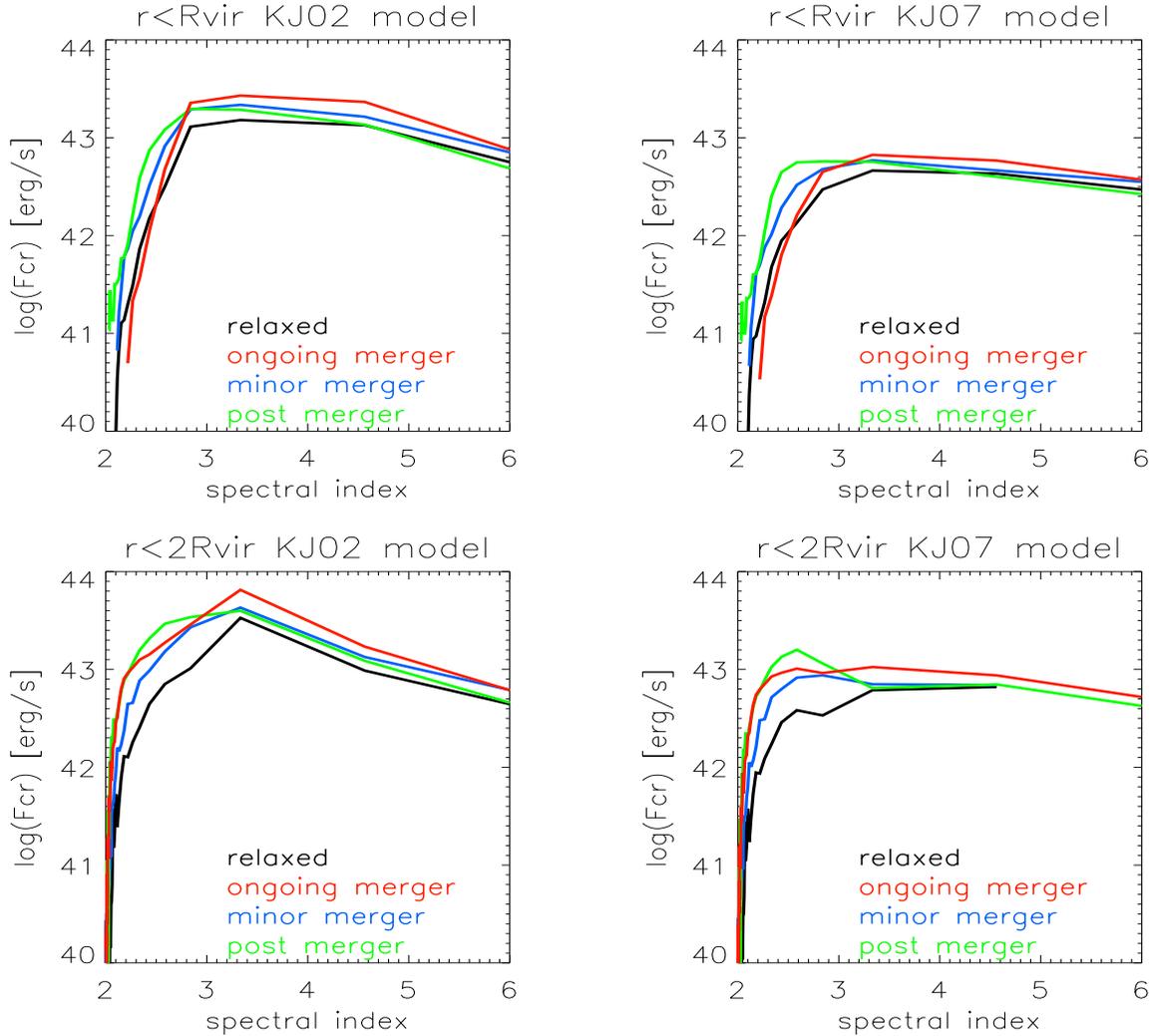}
\caption{Energy injection rate of CR for the 4 galaxy clusters
presented in the text, as a function of the 
spectral slope of the energy spectrum of injected protons, $N(p) \propto p^{-s}$. Estimates are shown for the KJ02 ({\it left}) and for the KJ07 model ({\it right}).
Here for simplicity we assume DSA at non--modified shocks (linear theory), 
in which case it
is $s = 2 (M^{2}+1)/(M^{2}-1)$ (e.g. Gabici \& Blasi 2003).}
\label{fig:elec_distrib}
\end{figure*}

\noindent
A comparison between our simulations and present limits
clearly requires a more detailed study.
However, 
a simple estimate of the spectral shape of CR injected in our
simulations (Fig.\ref{fig:elec_distrib}) suggests that 
the bulk of the CR energy
in clusters and cluster outskirts is associated with CR populations
with relatively steep spectra, in which case the limits from
radio observations become less stringent (Brunetti et al. 2007).
Obviously, an alternative possibility to
reconcile radio data and simulations 
would be that relatively flat
components of CR store a non negligible fraction of the energy of the
ICM, and that the magnetic field strengths in the ICM
are much smaller than that claimed from the analysis of present RM data;
future gamma ray observations with the FERMI Gamma Ray Telescope and
with Cherenkov arrays will clarify this possibility.

\bigskip

We thanks K. Dolag, D. Ryu, H. Kang, G. Tormen, 
L. Moscardini, B.O'Shea, M.Norman, R.Wagner, D.Collins, S. Skory, 
S. Giacintucci, R. Brunino and A.Bonadede for useful discussions and helps. 
We acknowledge partial support through grants ASI-INAF I/088/06/0 and
PRIN-INAF 2007, and the 
usage of computational resources under the CINECA-INAF agreement.


\bigskip

\bigskip

\appendix

{\bf APPENDIX}

\bigskip 

\section{The effect of spatial resolution.}
\label{sec:res_eff}

\begin{figure} 
\includegraphics[width=0.45\textwidth]{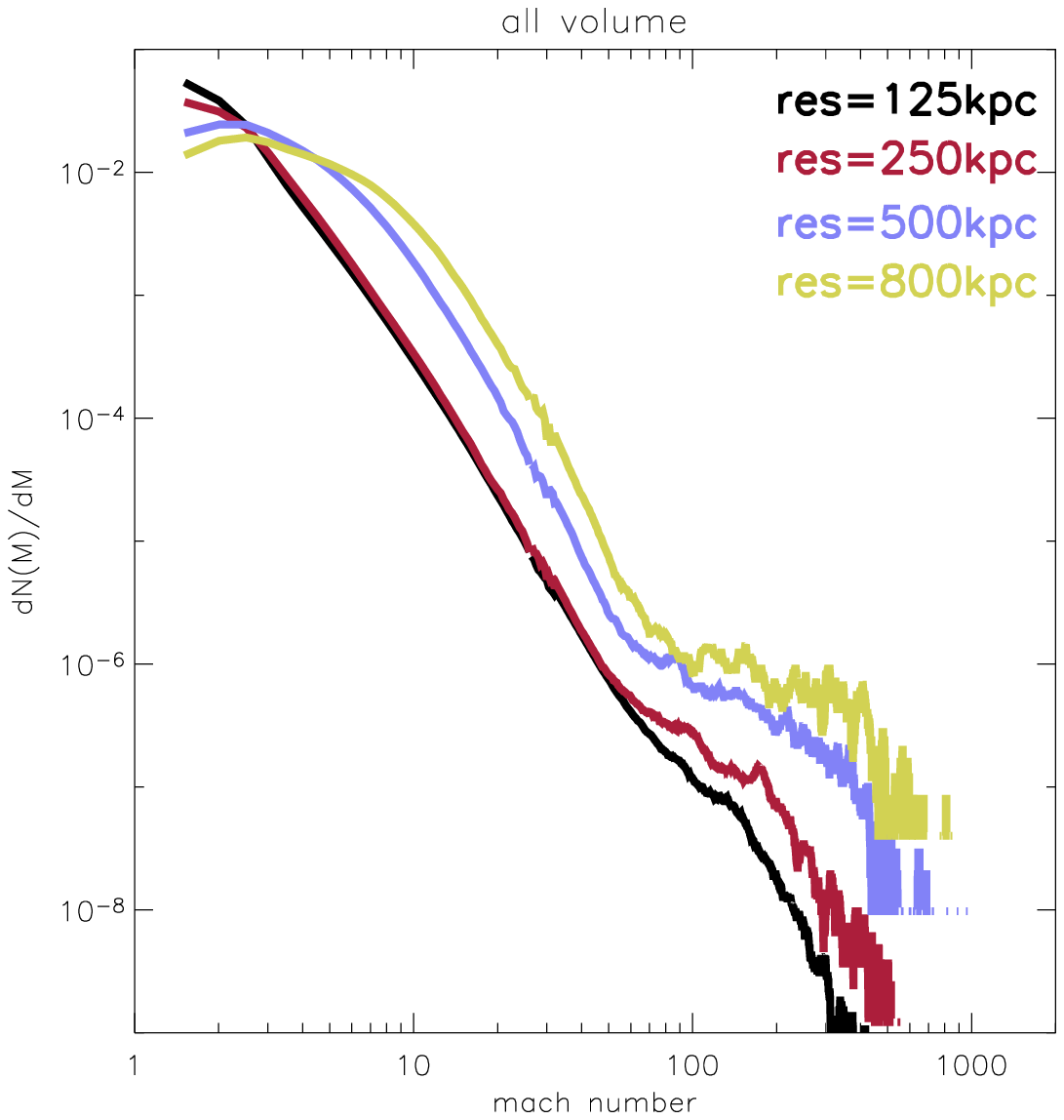}
\caption{Mach numbers distribution for the a $80Mpc$ box at 4 different numerical resolutions (i.e. $AD125$, $AD250$, $AD500$ and $AD800$).}
\label{fig:mach_no_res}
\end{figure}

We investigated the effect of resolution on the 
properties of detected shocks by re--simulating the
same initial conditions and cosmic volume of the $AD125$ simulations
at resolutions of $800kpc$, $500kpc$ and $250kpc$.

Even if most of the graphs and statistics presented in the paper
are done by using $n=1$ for the shock detecting scheme (see Sec.5.3)
and thus assuming that the best reconstruction of the shock discontinuity
is achieved by considering a jump of 2 cells between pre-shock and post-shock,
here we prefer to keep this jump smaller (i.e. $n=0$). This is in order
 minimize any confusion coming from the fact that in poorly resolved
runs shocks have sizes of typical cluster halos (i.e. for $n=1$ in the $AD800$
one would reconstruct shocks across $1.6Mpc$).
This is fair enough to reconstruct the trend with resolution within our simulations, and the comparison to the $n=1$ case can be recovered in Fig.14.

First we find that at all these resolutions 
Eq.~\ref{eq:fit2} provides a good fit to the density--temperature
distributions obtained with run 
time re--ionization and 
thus we use this relation to model the re-ionization in 
our post processing approach at all these resolutions.
We then analyze the outputs at $z=0$ and derive statistical properties
of shocks in the simulated volumes,
following the procedures given in the previous Sections.

The number distributions of shocked cells as a function of their Mach
number are given in Fig.~\ref{fig:mach_no_res} for the different
resolutions.
We find that the results converge at higher resolutions,
in particular
the shape of the distribution and integral number of shocked 
cells obtained with $125kpc$ 
and with $250kpc$ resolution are consistent within
$\approx 20$ percent.
A relevant point here is that the excess of shocks with high Mach number
found at low resolution is progressively reduced with increasing
resolution.

\begin{figure} 
\includegraphics[width=0.45\textwidth]{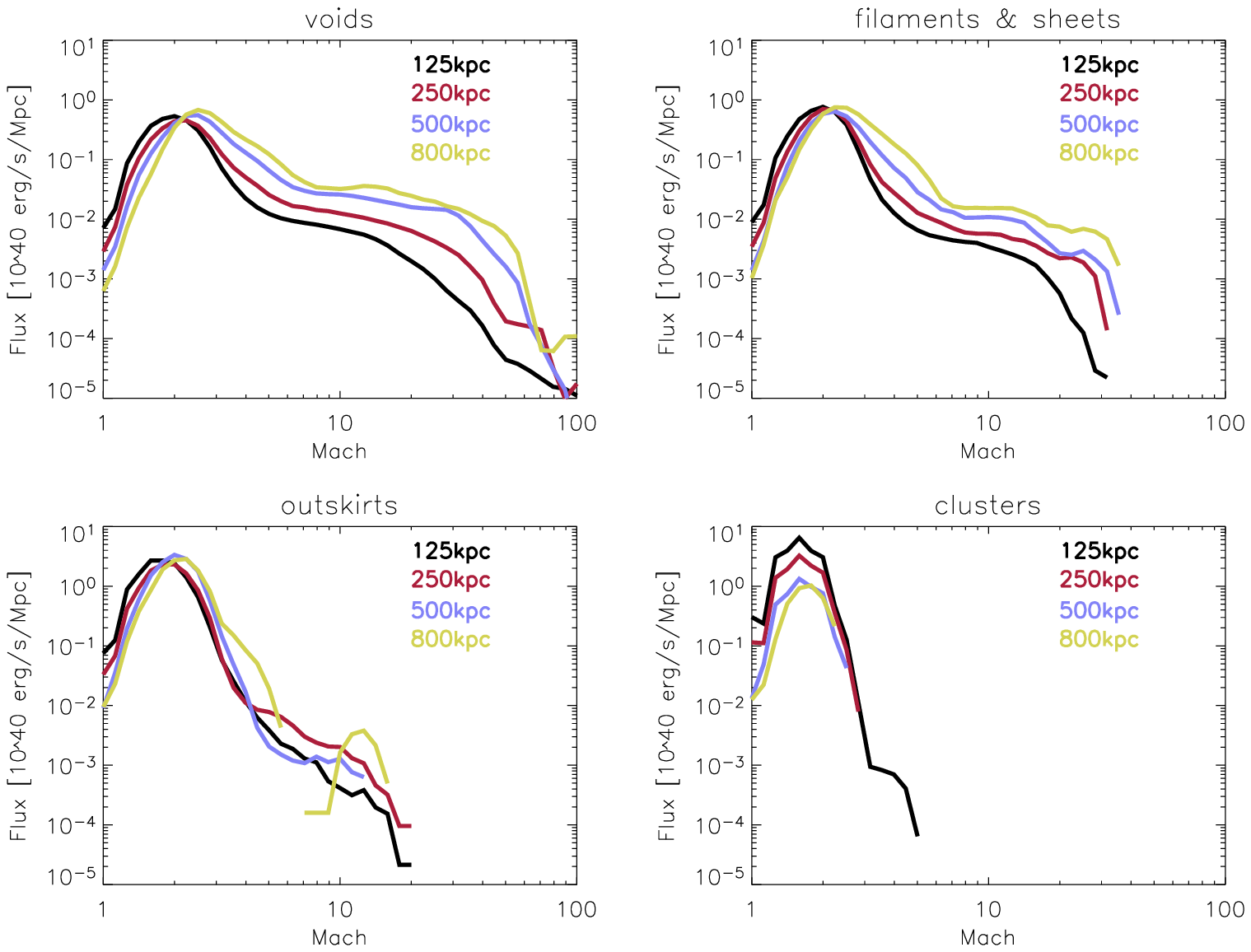}
\caption{Distribution of the thermalised energy flux in different over-density bins, for 
4 different numerical resolution.}
\label{fig:flux_dist_res}
\end{figure}

The case of the energy flux dissipated at shocks is
reported in Figures \ref{fig:flux_dist_res}.
This case is 
somewhat more problematic as it depends on the combination
of the properties of shocks with the local baryon over-density.
Despite the properties of shocks statistically converge with
resolution, 
the over-density in the simulated volume increases with
resolution and this causes the increase of the dissipated energy 
in the data simulated at higher resolutions.
Anyhow also in this case some level of convergence is obtained 
in line with previous studies 
(Ryu et al. 2003; Pfrommer et al. 2006a).
The most problematic case is that of galaxy clusters were 
indeed the dissipation of the energy at shocks increases by one order of
magnitude between lower and higher resolution data sets 
(this still increases by 
$\approx 1.5$ times between the $250$ and $125$ kpc data sets).

Despite this slow convergence with resolution, the value
of the ratio $f_{CR}/f_{th}$ is found to not significantly
change with resolution since the resolution affects the two quantities 
in a similar way.

\section{The effect of a variation of the $\sigma_{8}$ parameter.}
\label{sec:sigma8}

The value of the $\sigma_{8}$ parameter (the normalization in the power 
spectrum of
primordial over-density fluctuations) crucially affects the 
abundance of collapsed objects in the universe at a given epoch.
This value is presently not fully constrained: very recent CMB analysis  
give a relatively small value, $\sigma_{8} = 0.74$ (Spergel et al.2007),
with respect to that derived from previous CMB data--analysis
(Spergel et al.2003) and to the constraints from the observed 
abundance of galaxy clusters (e.g. Evrard et al.2007). 
In this Appendix we briefly discuss the 
effect of the $\sigma_{8}$ parameter
on the statistical properties of the shocks as measured
in our simulated data-sets (adopting as in the previous Section $n=0$ for the reconstruction of shocks).
We thus re simulated the {\it CO250} run with 
$\sigma_{8}=0.74$ ({\it S8250})
and applied all the procedures discussed in the previous Sections to
derive the properties of the shocks (note that the CO250 and S8250
simulations have run--time re-ionization).

Theoretically the population of shocks in a
universe with larger $\sigma_{8}$ is expected to
evolve faster as more power is associated with
the primordial over-density fluctuations.
Thus, at a fixed redshift, universes with larger $\sigma_{8}$
host more evolved structures, which are characterized by typically higher 
internal sound speeds at higher densities, and low temperatures in low
density regions.

The distribution of thermalised energy 
at shocks in the two simulations is given in
Fig.B1.
Although clearly modifying the value of $\sigma_{8}$ has some effect 
on the properties
of the shocks in the simulations, 
the net result is that, within the presently allowed region of
the values of the $\sigma_{8}$ parameter, no strong
difference in the properties of the shocks is found in  
simulations with different $\sigma_{8}$.
Indeed, globally we find that the energy dissipated at the present time
in the
{\it S8250} simulation is $\approx 2$ times smaller than
that in the {\it CO250} simulation, and the distribution with Mach
number of the dissipated energy in under-dense regions
is found to become somewhat   
slightly flatter with decreasing $\sigma_{8}$.

\begin{figure}
\includegraphics[width=0.45\textwidth]{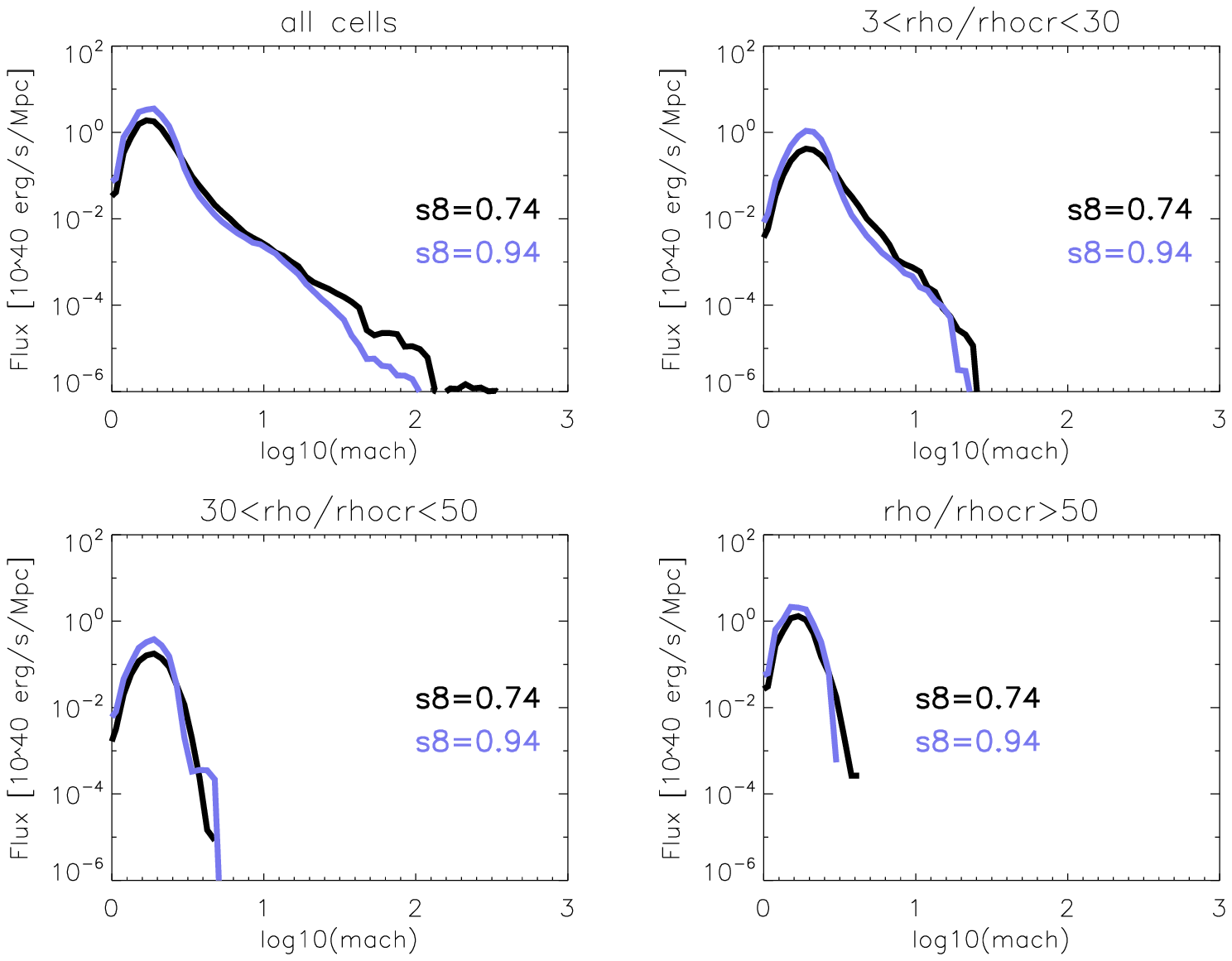}
\caption{Effect of the variation of the $\sigma_{8}$ parameter on the distribution of thermalised
flux, in different over-density bins.}
\label{fig:fl_s8}
\end{figure}

\end{document}